\documentclass[12pt,a4paper,final]{article}
\usepackage[margin=25mm]{geometry}
\usepackage[utf8]{inputenc}
\usepackage{graphicx}
\usepackage{caption}
\usepackage{subcaption}

\usepackage{longtable}
\usepackage{amsmath,amsfonts,amssymb,bef_alex,%
            ,bm,relsize,tikz}
\renewcommand{\ti}{{\times}}

\numberwithin{equation}{section}
\numberwithin{figure}{section}
\numberwithin{table}{section}

\usepackage[notref,notcite,color]{showkeys}

\usepackage{pifont}

\newtheorem{fact}[theorem]{Fact}
\newcommand{\CO}[1]{\underline{#1}} % correct digits
\newcommand{\IG}[1]{}    % ignore digits
\newcommand{\DLL}{\Lambda^*}   % dual lattice
\renewcommand{\ti}{{\times}}
\renewcommand{\dot}[1]{\overset{\text{\LARGE.}}{#1}}

\newcommand{\Ew}{\mafo{Ew}}

\newcommand{\mathOP}[1]{\mathop{\mathrm{#1}}} 
\usepackage{booktabs} % For \toprule, \midrule and \bottomrule

\begin{document}

\title{Symmetries in TEM imaging of crystals with strain%
\thanks{Partially supported by DFG through MATH+ (EXC-2046/1,
      project ID: 390685689) subproj.~EF3-1}} 

\author{Thomas Koprucki\thanks{Weierstraß-Institut f\"ur Angewandte
        Analysis und Stochastik, Mohrenstr.\,39, 10117 Berlin, Germany.},  
 Anieza Maltsi\footnotemark[2], and 
 Alexander Mielke\footnotemark[2]\ \thanks{Humboldt-Universit\"at zu Berlin, 
          Institut f\"ur Mathematik, Rudower Chaussee 25, 12489 Berlin
          (Adlershof), Germany.}  
}
 
%\date{Submitted: 11. Mai 2022}
\date{}

\maketitle

\begin{abstract} 
TEM images of strained crystals often exhibit symmetries, 
the source of which is not always clear. 
To understand these symmetries we distinguish between symmetries 
that occur from the imaging process itself and symmetries 
of the inclusion that might affect the image.
For the imaging process we prove mathematically that 
the intensities are invariant under specific transformations.
A combination of these invariances with specific properties 
of the strain profile can then explain symmetries 
observed in TEM images. 
We demonstrate our approach to the study of symmetries 
in TEM images using selected examples in the field 
of semiconductor nanostructures such as quantum wells 
and quantum dots.

\medskip

 \smallskip

\end{abstract}

\section{Introduction}
\label{se:intro}

 In transmission-electron microscopy (TEM) it is the main goal to extract
information on the specimen from the generated TEM images. This is particularly
used for detecting shapes, sizes, and composition of defects or inclusions like quantum wells
and quantum dots in a larger specimen consisting of a regular crystalline material. 
However, there is no direct way to infer the inclusion properties from
the TEM image. Hence, a commonly taken approach is to simulate the TEM imaging
process with inclusions being described by parametrized data. Then, the comparison with
experimental pictures can be used to fit the chosen parameters and deduce the
desired data of the experimental inclusions. 

A main feature in this process are symmetries for two reasons; first the
inclusions may have certain symmetries and second the TEM images may display symmetries that are 
related but not identical. 
The latter arises from the fact that the
experimental setup may have its own intrinsic symmetry properties.
In the present
work we want to analyze these symmetries and 
explain why sometimes TEM images
look more symmetric than the inclusion under investigation, 
or as the Curie's principle is stated 
in \cite{CasIsm2016CP} (a3): \emph{the effect is more symmetric than its cause}.   

The interest in TEM image symmetries dates back to the 1960'-70's, cf.\ \cite{HowWhe61DCEM,ISW74},
with the work focused mainly on the Reciprocity Theorem.
It states that \emph{the amplitude at a point B of a wave
originating from a source at point A and scattered by a potential $V$ is equal to the scattered amplitude at point A originating 
from the same source at B}. 
Many papers have been written for alternative proofs of this theorem, cf. \cite{BFTTM64,PT68,Moodie72,QG89}, as well as applications of it in the interpretation of TEM images, e.g.\ in connection with imaging of dislocations, cf.\ \cite{FTKKU72,HWM62,Katerbau80}.

While some of our results can also be deduced from the reciprocity theorem, like midplane reflection, 
there are more symmetries in the imaging process which can be proven mathematically by assuming the column approximation and focusing on the  Darwin--Howie--Whelan (DHW) equations \cite{Darw14TXRR12,HowWhe61DCEM}. Combining the symmetry properties of the imaging process with symmetry properties of the inclusion explains extra symmetries observed in TEM images of strained crystals.

The  Darwin--Howie--Whelan (DHW) equations, which
are often simply called Howie--Whelan equations (cf.\ \cite[Sec.\,2.3.2]{Jame90APTH} or
\cite[Sec.\,6.3]{Kirk20ACEM}), describe the propagation of electron beams
through crystals and can be applied to semiconductor nanostructures, see
\cite{Degr03ICTE, Pascal2018, MNBL19DDEG, MNSTK20NSTI}.
While these equations
are typically formulated for infinitely many beams in the dual lattice $\DLL$, for all
practical purposes it is sufficient to use only a few important beams, because
at high energy and for thin specimens only very few beams are excited by scattering
of the incoming beam. A mathematical analysis of the corresponding beam
selection is given in \cite{KMM21}, but this theoretical work is restricted to
perfect crystals without inclusions. Here we stay with finitely many
beams, i.e.\ with so-called $m$-beam models with wave vectors $\bfg \in
\DLL_m$, but generalize the analysis to
crystals with inclusions. The main assumption is however that the
crystallographic lattice stays approximately intact and can be modeled as a
strained crystal where the positions of the lattice points undergo a
displacement $\bfu(\bfr)$. Then, the 
DHW equation for strained crystals reads \begin{align}
    \label{eq:IntrDHW}
     \frac{\mathrm d}{\mathrm d z} \varphi_\mathbf{g}(z) 
     &= \mathrm{i} \pi \Big(2s_\mathbf{g} + (\mathbf{g}\cdot \frac{\mathrm d}{\mathrm d
     z} \mathbf{u}(x,y,z))\Big)\varphi_\mathbf{g}(z)+ \fr{\ii
     \pi}{\rho_\mathbf{g}}\sum_{\mathbf{h}\in \Lambda^*_m}
     U_{\mathbf{g-h}}\varphi_\mathbf{h}(z) \quad \text{for }\bfg\in \DLL_m.
\end{align}
Here $\psi_\bfg$ denotes the wave function of the beam associated with $\bfg\in
\DLL_m$, where $\bfg=\mathbf{0}$ denotes the incoming beam. The vertical coordinate $z\in
[0,z_*]$ gives the depth inside the specimen ($z=0$ entry plane and $z=z_*$
exit plane), whereas the horizontal coordinates $(x,y)$ are fixed and
correspond to the image pixel, see Figure \ref{fig:column_approx}. 

After a minor transformation the above system will take the vectorial form  
\begin{equation}
  \label{eq:Intr.m-beam}
   \dot \phi :=\frac\rmd{\rmd z} \phi = \ii\, \big(V+ \Sigma + F(z)\big)\,\phi
  \quad \text{and} 
\quad \phi(0)= \sqrt{\rho_\mathbf{0}}\,e_{\mathbf{0}} \in \C^m,
\end{equation}
where $\phi=(\phi_\bfg)_{\bfg\in \DLL_m} \in \C^m$ contains the relevant wave
functions. The Hermitian matrix $V$ corresponds to the electrostatic
interaction potential, the diagonal matrix $\Sigma= \mafo{diag}(s_\bfg)$
contains the so-called excitation errors, and $F(z)= \mafo{diag}(\bfg\cdot
\frac{\rmd}{\rmd z}\bfu(x,y,z)) \in \R^{d \ti d}$ contains the projections of the strains to
the individual wave vectors $\bfg\in \DLL_m$. We will call $F$ the \emph{strain profile}.

Image symmetries are now easily understood if changing the image pixel
$(x,y)$ to another pixel $(\wt x,\wt y)$ having the same strains throughout the
whole thickness, i.e.\ $\bfu(x,y,z)=\bfu(\wt x,\wt y, z)$ for all $z\in
[0,z_*]$, which implies $F(z)=\wt F(z)$. Such a situation is related to a
symmetry of the inclusion generating a symmetric strain field. 
As we will see, additional symmetries may occur in \eqref{eq:Intr.m-beam} 
in three distinct cases: 
\begin{enumerate}
\item 
  if $F(z)$ is replaced by $-F$, a so-called sign change;
\item
  if $F$ is reflected at the midplane $z=z_*/2$, i.e.\ $F(z)$ is replaced 
  by $F(z_*{-}z)$;
\item
 if $\Sigma$ is replaced by $-\Sigma$. 
\end{enumerate} 
The latter symmetry is relevant when a series of images are done while varying
the excitation error $s_\bfg$ along the series. 

These symmetries are observed experimentally (cf. \cite{MNBL19DDEG}) but occur for
the ODE system \eqref{eq:Intr.m-beam} only under additional
conditions. 
Typically the symmetries are exact only for the case of the
two-beam model with $\DLL_2=\{\mathbf{0},\mathbf{g'}\}$. Nevertheless, the
symmetries are approximately true in $m$-beam models if the intensities of the two
strong beams (bright field and dark field intensities) are much higher than those of
the weak beams. 

The structure of our paper is as follows: In Section \ref{su:TEMDHW} we provide
the background of TEM imaging, its numerical simulation via the DHW equations,
and the modeling of the influence of the strain. In Section
\ref{se:SymmetryTEM} we discuss all issues concerning symmetries in TEM imaging
by considering well-chosen examples. In particular, we highlight the relevance
of the symmetries for the detection of shapes of inclusions. The mathematical
rigorous treatment of the symmetries for the $m$-beam model
\eqref{eq:Intr.m-beam} is given in Section \ref{se:Symmetries}, where the
notion of weak and strong symmetries is introduced to provide a coherent
structure of the symmetry properties, which also reveals why the two-beam case is different from the $m$-beam case with $m>2$.

\section{TEM image formation and DHW equation}
\label{su:TEMDHW}

\begin{figure}
   \centering
    \includegraphics[width=0.9\textwidth]{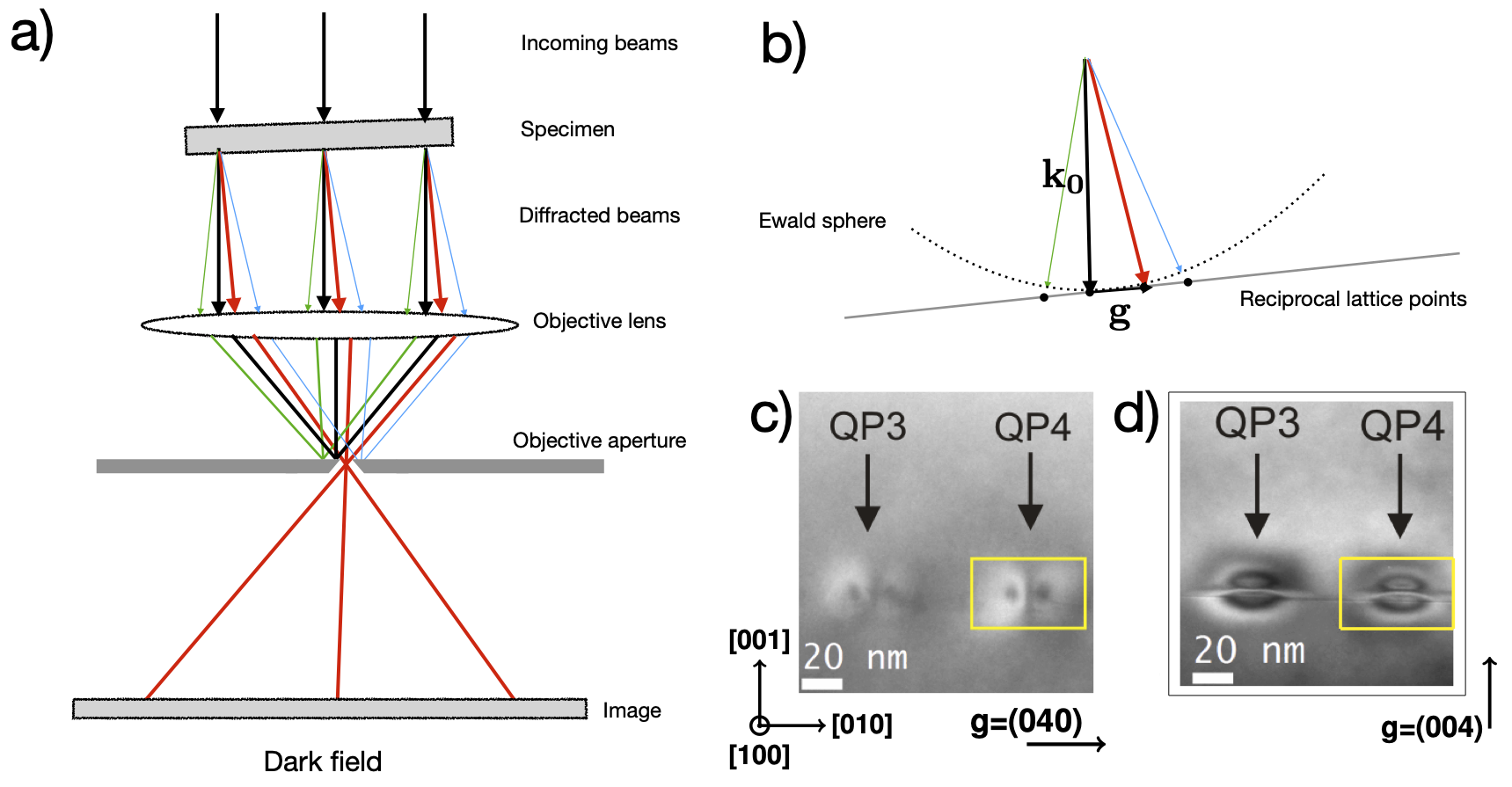}
     \caption{Image formation in TEM: (a) ray paths in TEM for  
     dark field imaging, where the objective aperture allows only
     selected diffracted beams (red) to pass to the detector.
     The incoming beam with wave vector $\mathbf{k_0}$ enters the
     specimen, is partially transmitted, and generates beams with
     nearby wave vectors $\mathbf{k_0 +g}$. 
     The red beam fulfills strong beam conditions on the Ewald
     sphere (b). Experimental TEM images of quantum dots for different choices of $\mathbf{g}$ (c) and (d). The whole figure is adapted from \cite[Fig.~1 and Fig.~2]{MNSTK20NSTI} used under CC-BY.
      }
     \label{fig:image_formation}
\end{figure}

In transmission electron microscopy electron beams are
transmitted through the specimen to create an image. 
A parallel electron beam illuminates the specimen. 
As specimen crystalline materials  
with a thickness of few hundred nanometers are considered.
Due to the periodic structure of the crystal
the electron beams are diffracted in discrete directions. 
The diffracted beams leaving the exit plane of the 
specimen are focused again by the objective. 
Then, with the objective aperture, the set of beams forming
the image can be reduced. 
This way specific beams can be chosen to create the image.
If the image that is created includes the 
undiffracted beam it is called \emph{bright field} image,
otherwise it is a \emph{dark field} image.
The ray path within the microscope for the creation of 
a dark field TEM image is illustrated in Figure~\ref{fig:image_formation} (a).

A crystal is a periodic structure created by the 
repetition of the unit cell across the directions 
of the direct lattice $\Lambda \subset \R^3$.
The reciprocal lattice $\DLL$ is the dual lattice of $\Lambda$ defined as
$
  \DLL:= \bigset{\mathbf{g} \in \R^3}{ \mathbf{g}\cdot \mathbf{r} \in \Z \text{ for all
    } \mathbf{r}\in \Lambda }. 
$
 The discrete directions that the beams are diffracted are
 given by Bragg's law \cite{Bra13} (also known as Laue conditions): 
 For an incoming beam with wavevector $\mathbf{k_0}$ a
 diffracted beam $\mathbf{k'}$ may occur if the condition
 $\mathbf{k'}= \mathbf{k_0} + \mathbf{g}$ is satisfied,
 where $\mathbf{g} \in \DLL$. 
 For elastic scattering the energy of the waves is conserved, meaning that the two vectors have the same length. 
 This implies that the wave vectors $\mathbf{k_0}$ and $\mathbf{k'}$
 have to lie on the surface of a sphere, known as the Ewald sphere \cite{Ewal21BOEG} and defined as $ \bbS_\Ew := \bigset{ \mathbf{g} \in \R^3 \:}{ \:|\mathbf{k_0}|^2 -
|\mathbf{k_0}{+}\mathbf{g}|^2 =0}.
$
When a reciprocal lattice point $\mathbf{g}$ falls on the Ewald sphere 
the Bragg condition is satisfied and a diffracted beam in the direction $\mathbf{k_0}+\mathbf{g}$ occurs, see Figure~\ref{fig:image_formation} (b).
However diffraction can occur even if the condition is not exactly satisfied. The deviation from Bragg's condition is expressed through the excitation error $s_\mathbf{g}$ defined as
 \begin{equation} \label{excerr}
       s_{\mathbf{g}} = -\fr{\mathbf{g} \cdot (2\mathbf{k_0} {+} \mathbf{g})}{2 | \mathbf{k_0} {+} \mathbf{g}| \mathrm{cos} \alpha} =
        \frac{|\mathbf{k_0}|^2- |\mathbf{k_0}{+}\mathbf{g}|^2}{ 2(\mathbf{k_0}{+}\mathbf{g}) \cdot \bfnu},
       \end{equation}
where $\alpha$ is the angle between the vector $\mathbf{k_0} + \mathbf{g}$ and the foil normal $\bfnu$. 
The excitation errors are parameters that can easily be controlled by the experimental conditions, like the tilting of the sample. For two-beam approximation we choose in addition to the incoming beam  $\mathbf{g}=\mathbf{0}$ one single reciprocal lattice vector $\mathbf{g'}\neq \mathbf{0}$ satisfying the so called \emph{strong beam conditions}, i.e. lies exactly on the Ewald sphere $s_\mathbf{g'}=0$. In this situation the two beams $\mathbf{g}=\mathbf{0}$ and $\mathbf{g}= \mathbf{g'}$ both are strongly excited because of $s_\mathbf{g}=0$.
Different choices can give rise to different image contrasts, as seen in Figures~\ref{fig:image_formation} (c) and (d).

\subsection{Multi-beam approach and DHW equations}

The electron propagation is described by the relativistic
Schr\"odinger equation, which is a 3D problem. 
However, for computational reasons, the 3D problem is often reduced 
to a 2D family of 1D problems using the so-called \emph{column approximation}. 
TEM uses fast electrons with acceleration voltages in the range
200-400 keV. This means that the angle between the diffracted and
undiffracted beam is very small.
For thin specimen (thickness in the range 100-200 nm) we can apply the
the column approximation, which states that an incoming beam will 
not leave a column centered at the entrance point. 
The width of the column defines the spatial resolution
and is typically in range of size of a unit cell, 
e.g. $\approx 0.5-1 \mathrm{nm}$.
It also assumes that electrons are not scattered in neighboring
columns and the propagation can be computed independently, by 
solving the equations for each column in turn. 

\begin{figure}
   \centering
    \includegraphics[width=\textwidth]{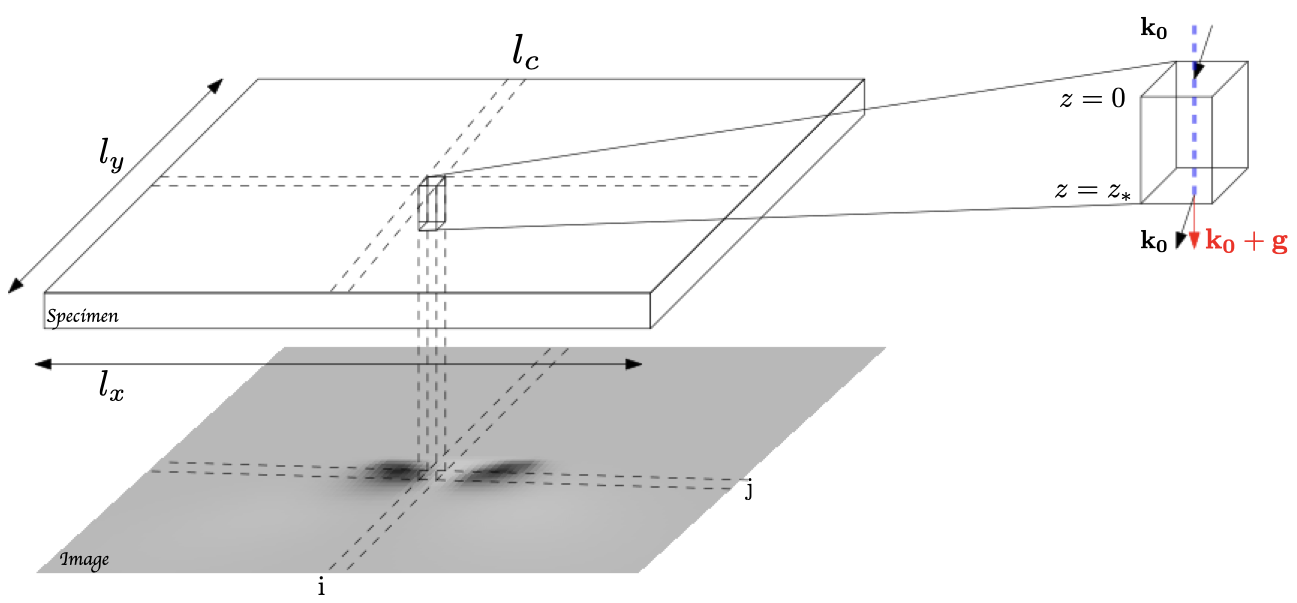}

     \caption{Column approximation: An incoming beam is assumed not to leave a column centered at the entering point. For this column the intensity corresponding to the $(i,j)$ pixel is obtained by propagating the beam along a line scan (blue line) in $z$-direction at position $(x_i, y_j)$ to the exit plane.  
     }
     \label{fig:column_approx}
\end{figure}

We divide a rectangular specimen into squares of edge length
$l_c$ defining the columns $(i,j), i=1,\dots N_x, j=1,\dots N_y$ 
centered around the positions $(x_i, y_j) \in \R^2$, 
see Figure \ref{fig:column_approx}. 
To obtain the simulated TEM image, for every pixel $(i,j)$ the
intensity has to be calculated by solving the dynamical diffraction
equations for that column. We decomposed the spatial variable 
$\mathbf{r} = (x, y, z)$ into the transversal part $(x,y)$ orthogonal 
to the thickness variable $z \in [0, z^*]$ , see Figure \ref{fig:column_approx}.
The propagation of the electron beam along the column 
is obtained by solving the \emph{Darwin-Howie-Whelan equations} numerically, like in pyTEM software \cite{Nier19pyTEM}, which in the case of a perfect crystal are:
\begin{align}
  \label{eq:I.DHW}
&  \frac{\mathrm d}{\mathrm d z} \psi_\mathbf{g}(z) = \mathrm{i} \pi \Big(2s_\mathbf{g}
  \psi_\mathbf{g}(z) + \fr{1}{\rho_\mathbf{g}}\sum_{\mathbf{h}\in \Lambda^*} U_{\mathbf{g-h}}\psi_\mathbf{h}(z) \Big), \quad
  \psi_\mathbf{g}(0)=\delta_{\mathbf{0},\mathbf{g}}, \quad \text{for } \mathbf{g}\in \DLL\\
& \text{where } \ \rho_\mathbf{g} = (\mathbf{k_0{+}g})\cdot \mathbf{\bfnu}, \nonumber
\end{align}
 where $s_\mathbf{g}$ are the excitation errors given in \eqref{excerr} and $U_\mathbf{g}$ are the Fourier coefficients of the periodic electrostatic lattice potential of the crystal. As the dual lattice $\DLL$ contains infinitely many points, \eqref{eq:I.DHW}  
is an initial value problem for an infinite system of first order ordinary differential equations describing the
propagation of the  electron beam through the specimen 
from the entry plane $z=0$ to the exit plane $z=z_*$. 

However, in experiments the setup is done in such a way that the incoming beam, which will always be given by $\mathbf{g}=\mathbf{0} \in \Lambda^*_0$, is diffracted in a few directions $\mathbf{k_0{+}g}$ for $\mathbf{g}$ lying in a small subset $ \Lambda^*_m $ of $ \DLL$, where $m$ is used to indicate the number of elements in $\Lambda^*_m $. Replacing $\DLL$ in \eqref{eq:I.DHW} by $ \Lambda^*_m$, we arrive at an $m$-beam model, which is an ODE for the vector $\big( \psi_\mathbf{g})_{\mathbf{g} \in \DLL_m} \in \C^m$.
Of special importance will be the 
\[
\text{\emph{two-beam model} with } \DLL_2=\{\mathbf{0},\mathbf{g'}\}.
\]
which is widely used. From now on we will denote by $\mathbf{g'}$ the diffracted beam in the two beam approximation and by $\bfg_\mafo{ap}$ the beam chosen by the objective aperture. For a bright field image we have $\bfg_\mafo{ap} = \mathbf{0}$ and for a dark field image under two beam approximation we have  $\bfg_\mafo{ap} = \mathbf{g'}$.

The problem of finding good subsets $\DLL_m$, which is the so-called beam selection problem, is discussed from a mathematical point of view in \cite{KMM21}. There it is argued that the infinite dimensional problem for $\mathbf{g} \in \DLL$ is even ill-posed, and it is shown that under typical assumptions the intensities $|\psi_\mathbf{g}(z) |^2$ decay exponentially like $\ee^{-\alpha | \mathbf{g}| }$. With this and further energetic considerations based on the Ewald sphere it was possible to derive rigorous error estimates to justify typical beam selection schemes, like the two-beam approximation or the systematic-row approximation.

\subsection{Influence of defects and strain}

\begin{figure}
    \centering
    \includegraphics[width=0.8\textwidth]{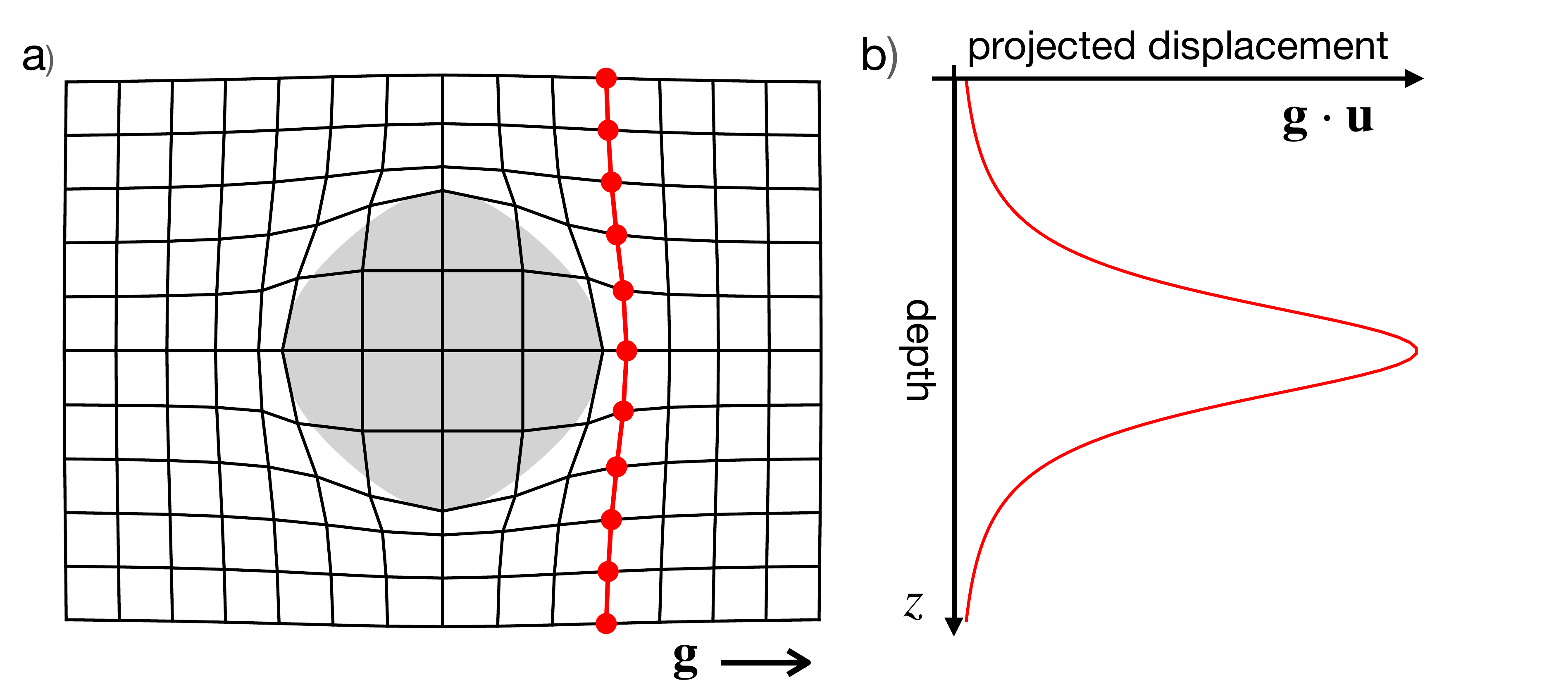}
    \caption{Crystal lattice with spherical inclusion:  a) deformation of the lattice b) variation of the projection of the displacement $\mathbf{u}$ on the $\mathbf{g}$ vector along the line scan in z-direction (red) in the crystal. }
        \label{fig:deformed_lattice}
\end{figure}

TEM imaging is widely used for the study of defects in crystalline materials, see \cite{Pascal2018,WuSch19TEMD,SchSta93,ZhuDeg20EBSD}.
Defects are perturbations of the crystal symmetry, in the sense that the atoms are displaced from their original position in the perfect crystal. If an atom was at position  $\mathbf{r}$, its new position will be 
$\mathbf{r'} = \mathbf{r}+ \mathbf{u}(\mathbf{r})$, where $\mathbf{u}(\mathbf{r})$ is 
the displacement field, see Figure \ref{fig:deformed_lattice}~a). 
As an elementary example for strained crystals we consider 
a spherical particle with radius $r_0$ and lattice parameter $a_p$ inside a matrix with lattice parameter $a_m$, as is done in \cite[Ch.8, p.479]{Degr03ICTE}. 
The displacement field is given by
\begin{equation} \label{eq:sphere_displacement}
\bfu(\bfr) = C(\delta) \frac{\big(\min\{|\bfr|,r_0\}\big)^3}{|\bfr|^3}\,\bfr
\end{equation}
where $C(\delta)$ is a constant that depends on the elastic properties of the isotropic matrix and $\delta$ the matrix misfit given by $\delta = (a_p - a_m)/a_m$.
 In this case the displacement inside the particle is proportional to $\bfr=(x,y,z)$, whereas outside it decays as $1/|\bfr|^2$, see Figure \ref{fig:deformed_lattice}. The displacement field $\bfu$ is only valid for small isotropic inclusions where the particle diameter is significantly smaller than one extinction distance. An example of such a case can be a spherical InAs quantum dot inside a GaAs matrix, see \cite{MNSTK20NSTI, NiermannPhD21}.

For small deformations the displacement will modify 
the Fourier coefficients of the potential in the DHW equations \eqref{eq:I.DHW} by a phase factor
\[
U_\mathbf{g} \rightarrow U_\mathbf{g} \ee^{-2\ii\pi \mathbf{g} \cdot \mathbf{u}(\mathbf{r})}.
\]
Using this and letting $\psi_\mathbf{g} = \varphi_\mathbf{g} \ee^{-\ii \mathbf{g} \cdot \mathbf{u}(\mathbf{r})}$ in \eqref{eq:I.DHW} we get the DHW equations for a strained crystal \cite[Ch.8]{Degr03ICTE}:
\begin{align}
    \label{eq:DHWdef}
     \frac{\mathrm d}{\mathrm d z} \varphi_\mathbf{g}(z) 
     &= \mathrm{i} \pi \Big(2s_\mathbf{g} + \frac{\mathrm d}{\mathrm d z}(\mathbf{g}\cdot \mathbf{u}(\mathbf{r}))\Big)\varphi_\mathbf{g}(z)+ \fr{\ii \pi}{\rho_\mathbf{g}}\sum_{\mathbf{h}\in \Lambda^*_m} U_{\mathbf{g-h}}\varphi_\mathbf{h}(z)
  \\
     \text{and } \varphi_\mathbf{g}(0)
     & =\delta_{\mathbf{0},\mathbf{g}} \quad \text{for } \mathbf{g}\in \DLL_m.  
\end{align}

To simulate a TEM image with defects the column approximation and the DHW equation as described above can still be used, but now for each horizontal position $(x_i,y_j)$, where $(i,j)$ denotes the image pixel, $\mathbf{u}(\mathbf{r}) $ in \eqref{eq:DHWdef} is evaluated as  $\mathbf{u}(z;x_i,y_j)$. If this is constant, then the defect will not be visible.
Another important fact for the imaging of defects is that the projection of the displacement to the reciprocal lattice vector $\mathbf{g}$ is what really matters, see Figure \ref{fig:deformed_lattice} b).
If $\bfg \cdot \bfu(\mathbf{r})$ is constant, then again the defect is not visible. 
 This means that by choosing different vectors $\bfg_\mafo{ap}$ we get different information about the defect.

\begin{figure}
    \centering
    \includegraphics[width=\textwidth]{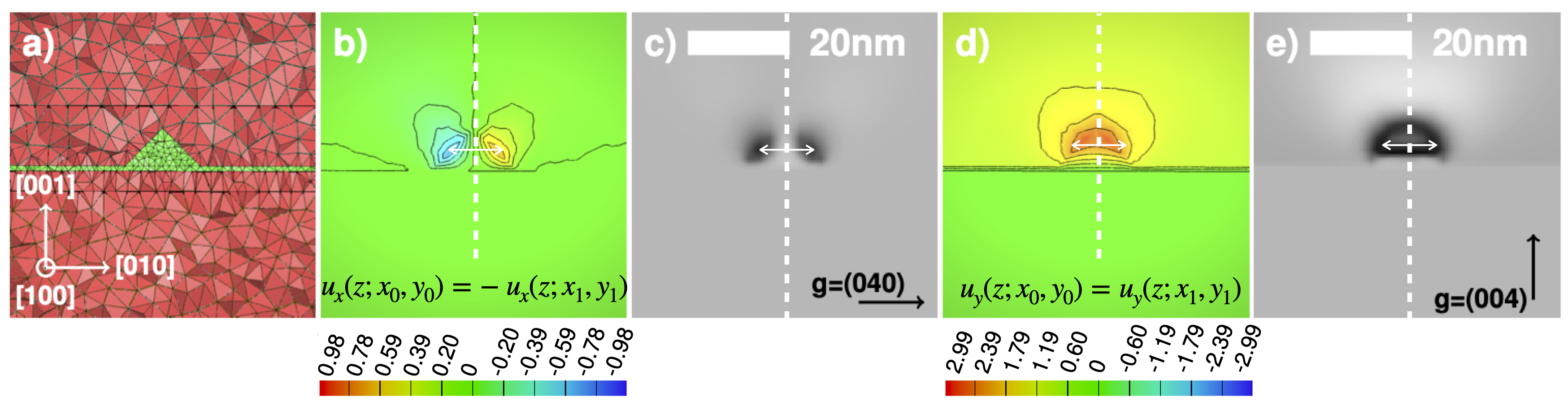}
    \caption{ Simulation of TEM images for pyramidal QD:
                 a) QD geometry indicating the crystallographic directions b) $u_{x}$ component and d) $u_{y}$ component of displacement field along a cross-section in the center of the structure. c) and e) corresponding TEM images for strong beam conditions as indicated by the direction of the chosen vectors $\bfg$. The images in the figure are adapted from \cite[Fig. 5]{MNSTK20NSTI} used under CC-BY. 
                 }
    \label{fig:sim_pyramid}
\end{figure}

Figure \ref{fig:sim_pyramid} illustrates this for a pyramidal quantum dot. Choosing $\bfg_\mafo{ap} = (040)$ will create a TEM image corresponding to the $u_x$ component of the displacement, as seen in Figure \ref{fig:sim_pyramid} b) and c). Changing to $\bfg_\mafo{ap} = (004)$ will give a TEM image corresponding to the $u_y$ component of the displacement, see Figure \ref{fig:sim_pyramid} d) and e). This sensitivity of TEM images to different components of the displacement field is important for the interpretation of images and can be used for the reconstruction or classification of the observed object.
In \cite{MNSTK20NSTI} this was used to compare quantum dots of four different geometries. It was observed that the projection to the vector $\bfg_\mafo{ap}= (004)$ would give better contrast, allowing one to distinguish between pyramidal or lense shaped dots, while in the $\bfg_\mafo{ap} = (040)$ direction all images would show a very similar coffee-bean contrast making it difficult to distinguish among the geometries.

\section{Symmetries in TEM images}
\label{se:SymmetryTEM}

In this section we study observed symmetries in TEM images 
of strained crystals  and discuss their interpretation.
To this purpose we introduce selected examples demonstrating 
different kinds of symmetries, e.g. images 
that are pixelwise symmetric, 
like c) and e) in Figure \ref{fig:sim_pyramid}. 
This kind of symmetry occurs when there is a sign change 
in the displacement component, see Section \ref{su:obs_sign_dis}, 
or when the inclusion is shifted from the center, 
see Section \ref{su:obs_shift}. 
In Section \ref{su:obs_sign_err} symmetries of a series of TEM images for varying excitation errors $s_\mathbf{g}$ 
and varying positions are discussed. 
These examples show the importance to distinguish 
different kind of symmetries that can occur and to 
examine which ones are connected to specific properties of the displacement or strain profile and which are independent of it. 

To understand the origin of these symmetries, we performed an analysis on the symmetry properties of solutions of the DHW equations. 
The main results are explained in 
Section \ref{su:SymmExplainDhW}, while the formal proofs are given in Section \ref{se:Symmetries}. 
This analysis revealed three important symmetry principles, stated in Section \ref{suu:PhysSymmProved}. 
By combining these principles with specific properties of the strain profiles we can explain all the observed symmetries 
introduced in Sections \ref{su:obs_sign_dis}-\ref{su:obs_sign_err}.
The capability of our approach to explain symmetries in
TEM images beyond these examples is demonstrated in 
Section \ref{suu:general_prof}, where the developed theory 
is applied to a more complex problem featuring 
general displacement profiles.

\subsection{Symmetry with respect to the sign of the displacement}
\label{su:obs_sign_dis}

In Figure \ref{fig:sim_pyramid} c) and e) we see two simulated TEM images for different choices of the vectors $\mathbf{g_{ap}}$. Each image is pixelwise symmetric, in the sense that for two different pixels $(x_0,y_0)$ and $(x_1,y_1)$ we have the same intensities:
$I_\mathbf{g_{ap}}(x_0,y_0)=|\varphi_\mathbf{g_{ap}}(z_*;x_0,y_0)|^2 = |\varphi_{\mathbf{g_{ap}}}(z_*;x_1,y_1)|^2= I_\mathbf{g_{ap}}(x_1,y_1)$. For image \ref{fig:sim_pyramid} e) this in not surprising since the  profile of the vertical component of the displacement  along the column related to pixel $(x_0,y_0)$  
is the same as the one for pixel $(x_1,y_1)$, namely $u_y(z;x_0,y_0) = u_y(z;x_1,y_1)$ for $z\in[0,z_*]$,
due to the symmetry of the pyramid. However, the pixelwise symmetry in image \ref{fig:sim_pyramid} c) is interesting: the profiles of the horizontal displacement component, which are responsible for the image contrast,  have opposite values  $u_x(z;x_0,y_0) = -u_x(z;x_1,y_1)$. 
This indicates that there might be some symmetry in TEM images with respect to the sign of the displacement. 

In \eqref{eq:DHWdef} we see that it is the product of the strain $\fr{d}{d z} \bfu$ with the reciprocal lattice vector $\mathbf{g}$ 
that enters the equations. 
This term will from now on be expressed as $F_\mathbf{g} (z) = \fr{d}{dz} (\mathbf{g}\cdot \bfu(z;x_i,y_j))$
and the influence of the strain to a $m$-beam system 
will be represented by the matrix-valued function $F(z) =\mathOP{diag} \big( F_\mathbf{g}\big)_{\bfg \in \DLL_m}$. 
So we want to know if the transformation $F(z) \leadsto -F(z)$ gives the same intensity.
If it does, then the question that arises is whether it is for a specific shape of the strain profile $F(z)$
or it is independent of it and applies to general strain profiles.
\subsection{Symmetry with respect to the center of the sample}
\label{su:obs_shift}
\begin{figure}
    \centering
    \includegraphics[width=\textwidth]{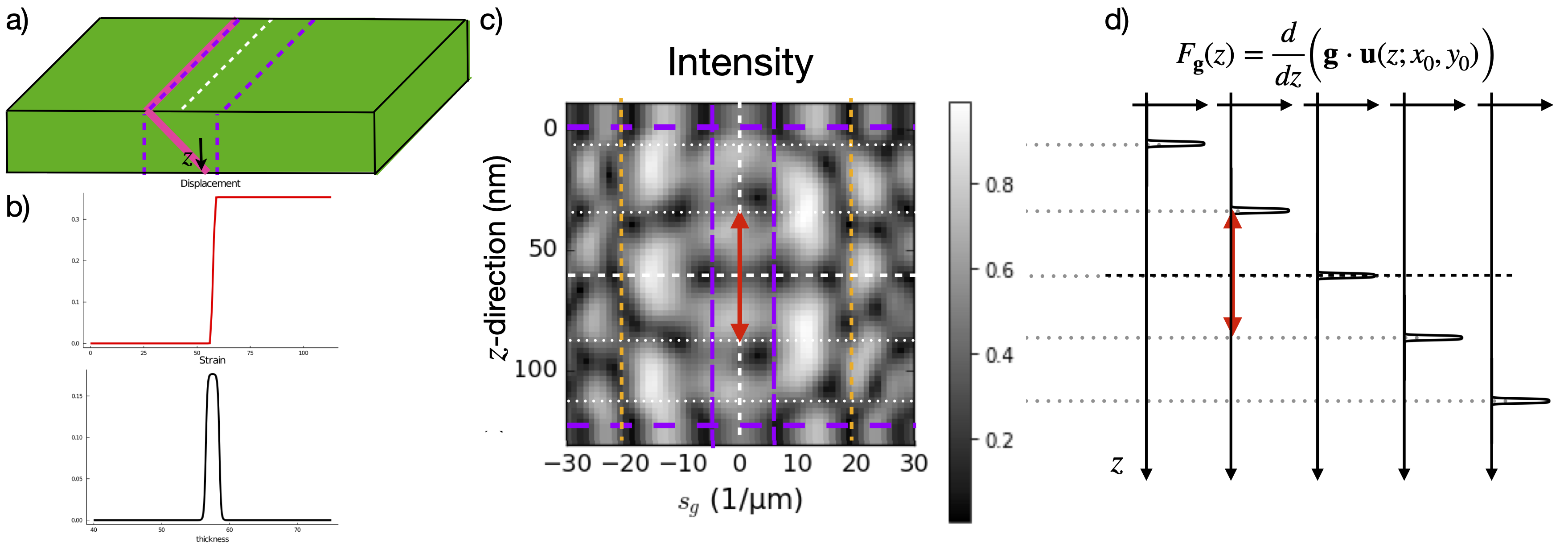}
    \caption{TEM imaging of inclined quantum well: a) illustration of a specimen for an inclined quantum well showing the $z$-direction and two line scans (purple dotted lines). b) The displacement (red) and strain (black) profiles projected to the reciprocal vector $\mathbf{g_{ap}}$. c)
    Intensity values for different positions and different excitation errors $s_\mathbf{g_{ap}}$ for a beam propagating in $z$ direction. Adapted from \cite[Fig.5]{MNBL19DDEG} used under CC-BY. d) Strain profile for the different positions corresponding to a shift of the strain across the $z$ direction. }
    \label{fig:quantum_well}
\end{figure}
\begin{figure}
    \centering
    \includegraphics[width=\textwidth]{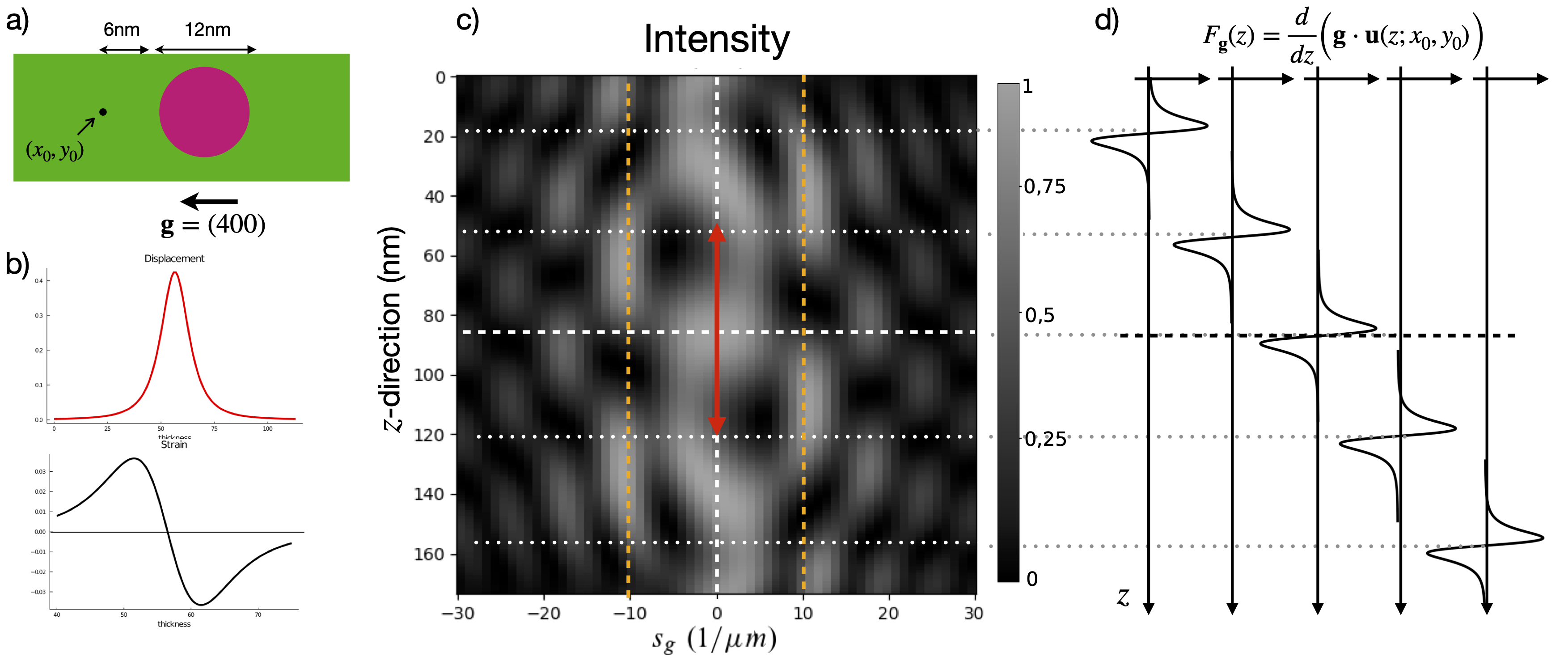}
    \caption{TEM imaging of spherical quantum dot: a) illustration of the specimen in $(x,y)$ projection showing the direction of the chosen beam $\mathbf{g_{ap}}$.
    b) Propagating the beam across $z$ for a chosen $(x_0,y_0)$ gives the intensity at the corresponding pixel. The displacement (red) and strain (black) profiles projected to the reciprocal vector $\mathbf{g_{ap}}$.
    c) Intensity values for different positions and different excitation errors $s_\mathbf{g_{ap}}$. Adapted from \cite[Fig. 5.13]{NiermannPhD21} used under CC-BY. 
    d) Strain profile for the different positions corresponding to a shift of the strain across the $z$ direction. }
    \label{fig:quantum_dot}
\end{figure}
Our next example is inspired by images provided in \cite{MNBL19DDEG}, where TEM imaging of an inclined strained semiconductor quantum well, like the one in Figure \ref{fig:quantum_well} a),  has been studied. A quantum well is a planar heterostructure consisting of a thin film, forming the quantum well, sandwiched between barrier material layers forming the matrix. 
Due to the lattice mismatch between the materials
the lattice of the quantum well is deformed. 
For pseudomorphically grown quantum wells with perfect interfaces it can be assumed that the displacement 
grows linearly within the quantum well region and has a constant value outside, resulting in a strain profile similar to an indicator function, see Figure \ref{fig:quantum_well} b).

The intensity values of the dark field for such a structure are shown in Figure \ref{fig:quantum_well} c), for different values of the excitation error and for different positions. Due to the incline angle between the planar interface and the imaging direction the different positions correspond to different depths of the quantum well as measured from the surface of the specimen, see \ref{fig:quantum_well} a).
An interesting first observation here is that the intensity seems to be symmetric with respect to a shift in the position from the center of the sample and for every excitation error $s_\mathbf{g_{ap}}$. A natural question that occurs is whether this shifting symmetry is a general property of TEM imaging. The answer is negative and this can be seen in Figure \ref{fig:quantum_dot} c) which shows the dark field intensities for a spherical quantum dot (Figure \ref{fig:quantum_dot} a) again for different excitation errors and different positions. Shifting the quantum dot from the center for an excitation error $s_\mathbf{g_{ap}} \neq 0$ does not give the same intensity. However, if we choose $s_\mathbf{g_{ap}}=0$ then we observe again a symmetry with respect to shifting. 

To analyze these observations we take a closer look into the shape of the strain in each case. For the quantum well in Figure \ref{fig:quantum_well} the strain profile is an even function (Figure \ref{fig:quantum_well} b)) while for the  quantum dot in Figure \ref{fig:quantum_dot} it is an odd function (Figure \ref{fig:quantum_dot} b)). The latter is  due to the symmetry of the sphere, cf. displacement field for spherical inclusion \eqref{eq:sphere_displacement}.
Shifting the inclusion would correspond to shifting the strain in both cases as seen in Figures \ref{fig:quantum_well} d) and  \ref{fig:quantum_dot} d), respectively. 
The questions to be answered here are i) what is special in the case $s_\mathbf{g_{ap}}=0$ that makes shifting a symmetry, ii) how does shifting an even or odd strain profile affect the intensities and iii) what happens for a general strain profile?

\subsection{Symmetry with respect to the sign of $s_\mathbf{g}$}
\label{su:obs_sign_err}

In the previous examples we considered pixelwise symmetry for one specific image. This was expressed as $I_\mathbf{g_{ap}}(x_0,y_0)= I_\mathbf{g_{ap}}(x_1,y_1)$.
In this section we talk about pixelwise symmetry between images. This means that if $I_\mathbf{g_{ap}}$ corresponds to the intensity of an image and  $\wt I_\mathbf{g_{ap}}$ to the intensity of another image then the two images are pixelwise symmetric if $I_\mathbf{g_{ap}}(x_i,y_j) = \wt I_\mathbf{g_{ap}}(x_i,y_j) $ for every pixel $(i,j)$.

\begin{figure}
    \centering
    \includegraphics[width=\textwidth]{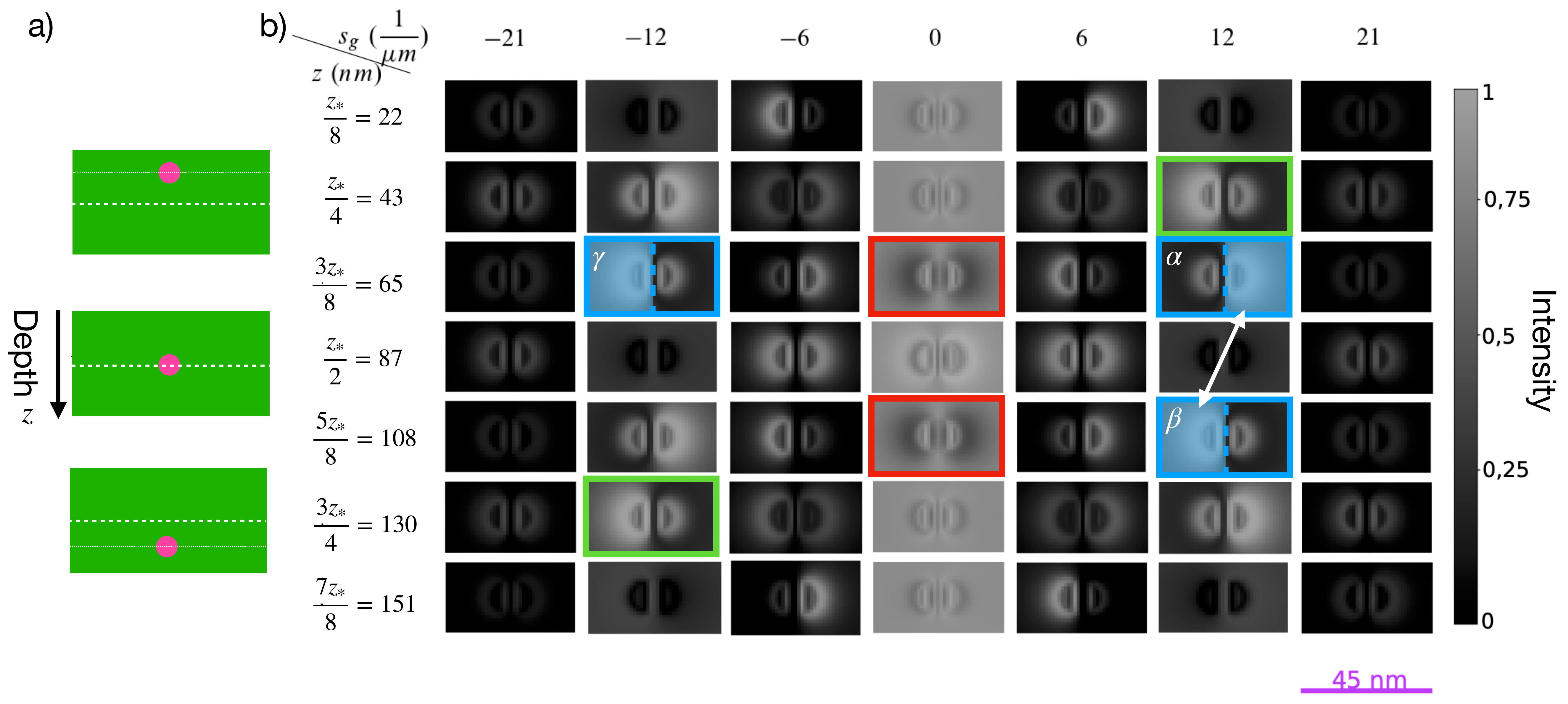}
    \caption{Series of TEM images for a spherical quantum dot:
    a) schematics of the position of quantum dot in the sample.
    b) simulated TEM images for different depths of the quantum dot and for different excitation errors $s_\mathbf{g_{ap}}$. For $s_\mathbf{g_{ap}}=0$ the TEM images show a pixelwise symmetry with respect to the center (red boxes). For $s_\mathbf{g_{ap}} \neq 0$ the TEM images are symmetric with respect to the center if in addition the sign of the excitation error is changed (green boxes). The images are mirrored to each other with respect to the center for the same excitation error ( $\alpha$ and $\beta$ blue boxes) or with respect to the sign of the excitation error for a fixed position ($\alpha$ and $\gamma$ blue boxes). Adapted from \cite[Fig.5.12]{NiermannPhD21} used under CC-BY. }
    \label{fig:qd_spherical_intensities}
\end{figure}

In Figure \ref{fig:qd_spherical_intensities} we have TEM images, adapted from \cite{NiermannPhD21}, 
of a spherical quantum dot at different positions and for different excitation errors.
The observations we made for shifting at the previous section apply here as well. Shifting the quantum dot for an excitation error $s_\mathbf{g_{ap}}=0$ creates images that are pixelwise symmetric with each other (red boxes) while for $s_\mathbf{g_{ap}} \neq 0$ they are not symmetric ($\alpha$ and $\beta$ blue boxes). Shifting for an $s_{\mathbf{g_{ap}}} \neq 0$ however seems to create mirrored images, in the sense that the image $\alpha$ is a mirrored version of image $\beta$ with respect to the symmetry axis of the sphere.

Interestingly though we see that if we shift the quantum dot from the center and additionally change the sign of the excitation error $s_\mathbf{g_{ap}}$ then the two images are pixelwise symmetric (green boxes or blue $\beta$ and $\gamma$ boxes). 
Again the question that arises here is whether these observations are connected to a specific property of the strain profile or is there a symmetry connected to shifting and sign change of $s_\mathbf{g_{ap}}$ that occurs for general profiles?  

\subsection{Symmetries explained via DHW equations}
\label{su:SymmExplainDhW}

To understand the symmetries in TEM images
we described above, we studied the properties 
of the beam propagation through the specimen
using the DHW equations. 
It turned out, that the intensity at the exit plane 
is invariant under specific transformations of  
the strain field. In the following we give an introduction 
to our approach and an overview of the different types 
of symmetries formally 
defined and proved in mathematically rigorous terms
in Section \ref{se:Symmetries}. We conclude the section with an explanation of the observed symmetries using the theory we developed.

\subsubsection{Transformation to Hermitian form}
\label{suu:TrafoHerm}

To begin with, it is essential to use the self-adjoint structure that is somehow hidden in the DHW equations. This can either be done as in \cite{KMM21}, where $\C^m$ is equipped with the scalar product $\big< a, b \big> = \sum_{\bfg\in \DLL_m} \rho_\bfg a_\bfg \ol{b}_\bfg$, or by the simple transformation 
\[
\phi_\bfg = \sqrt{\rho_\bfg}\: \varphi_\bfg \text{ for } \bfg \in \DLL_m,
\] 
which will be used in this paper. This has the advantage that $\C^m$ is equipped with the standard (complex) Euclidean scalar product, but the intensities take the form $I_\bfg(x_i,y_j)=|\varphi_\bfg(z_*;x_i,y_j)|^2 =  |\phi_\bfg(z_*;x_i,y_j)|^2/ \rho_\bfg $.

In terms of $\phi=\big( \phi_\bfg\big)_{\bfg \in \DLL_m} \in \C^m$, the system \eqref{eq:DHWdef} is rewritten in matrix form as follows:
\begin{subequations}
  \label{eq:m-beamHerm}
  \begin{equation}
  \label{eq:m-beam}
  \dot \phi :=\frac\rmd{\rmd z} \phi = \ii\, \big(A+ F(z)\big)\,\phi \quad \text{and}
\quad \phi(0)= \sqrt{\rho_\mathbf{0}}\,e_{\mathbf{0}} \in \C^m.  
\end{equation}
Subsequently, we will omit the normalizing factor $\sqrt{\rho_\mathbf{0}}$ in the initial condition $\phi(0)$, because it is not relevant in TEM imaging, where gray-scale pictures are created using relative intensities only.  
The system matrix $A=V+\Sigma $ and the influence $F(z)$ of the strain are given via 
\begin{equation}
  \label{eq:V.Sigma.F}
  V=\Big( \frac{ \pi U_{\bfg-\bfh} } 
  { \sqrt{ \rho_\bfg \rho_\bfh} } \Big)_{\bfg,\bfh\in \DLL_m}, 
\quad 
 \Sigma =\mathOP{diag}(2\pi s_\bfg)_{\bfg\in \DLL_m}, 
\quad  F(z) =\mathOP{diag} \big( \frac{\rmd}{\rmd z}(\bfg \cdot \bfu(z))\big)_{\bfg \in \DLL_m},
\end{equation}
\end{subequations}
where $V$ describes the interaction of the beams 
via the scattering potential and $\Sigma$ is related
to the excitation conditions.
As the Fourier coefficients of the scattering potential satisfy $ U_{-\bfg} = \ol U_\bfg$, we see that $V\in \C^{m\ti m}$ is indeed a Hermitian matrix,
while $\Sigma$ and $F(z)$ are real-valued diagonal matrices. 

What is important in TEM imaging is the intensity of the strongly excited beams at the exit plane $z=z_*$ and not all components of $\phi$. Our theory is developed in such a way that it focuses on the amplitude of the undiffracted beam, $|\phi_\mathbf{0}(z_*)|$, which corresponds to a bright-field image. The point is that this generates a potential reflection symmetry $z \leadsto z_*-z$, because the initial condition $\phi(0)=e_0$ and the exit measurement $\phi_\mathbf{0}(z_*)= \phi(z_*)\cdot e_{\mathbf{0}}$ use the same vector $e_{\mathbf{0}}$.  

Intensities of solutions for different choices of the pair $(A,F(z))$ are compared to see which replacements of $(A,F)$ by $(\wt A,\wt F(z))$ lead to the same (measurement) results. 
Such transformations are then called symmetries. Changes in $F(z)$ correspond to transformations in the strain, while changes in the matrix $A$ can correspond to 
transformations in the excitation errors (given by $\Sigma$) or the potential (given by $U$).

\subsubsection{Strong and weak symmetries}
\label{suu:PhysWeakStrong}

Two kind of symmetries are defined in Section \ref{def:Symmetries}: \emph{strong} and \emph{weak symmetries.} 
For \emph{strong symmetry} the intensity
of the beam along the whole column $[0,z_*]$
is invariant under the transformation
$(A,F) \rightarrow (\wt A, \wt F)$, that means
the corresponding solutions $\phi$ 
and $\wt \phi$ satisfy $|\phi_\mathbf{0}(z)| = |\wt \phi_\mathbf{0}(z)|$ 
for all $z \in [0,z_*]$. 
For \emph{weak symmetry} this invariance holds for the intensity of the beam at the exit plane only, namely 
$|\phi_\mathbf{0}(z_*)| = |\wt \phi_\mathbf{0}(z_*)|$. 
In TEM imaging this distinction is not visible since 
we only see the intensity at the exit plane. 
For the mathematical analysis 
however this distinction is highly relevant because of the different underlying mechanisms. Of course, any composition of weak and strong symmetries provides a weak symmetry again. 

\begin{figure}
    \centering
       \begin{subfigure}{0.4\textwidth}
         \includegraphics[width=\textwidth]{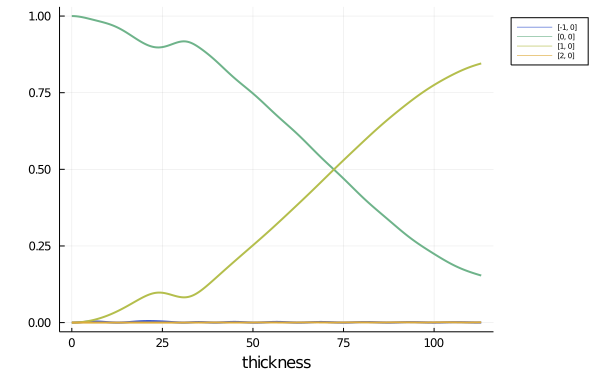}
          \caption{System $(A,F(z))$}
       \end{subfigure}
      \hfill
       \begin{subfigure}{0.4\textwidth}
         \includegraphics[width=\textwidth]{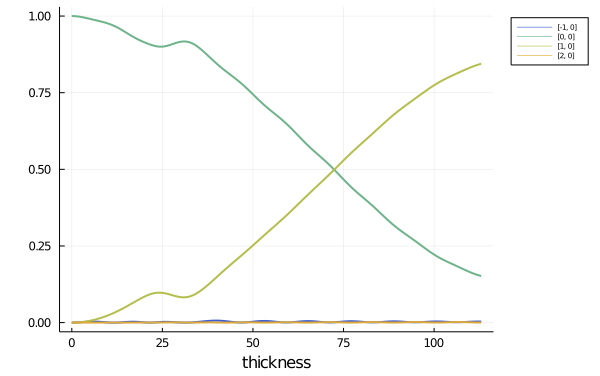}
          \caption{System $(A,-F(z))$}
       \end{subfigure}
      
       \begin{subfigure}{0.4\textwidth}
         \includegraphics[width=\textwidth]{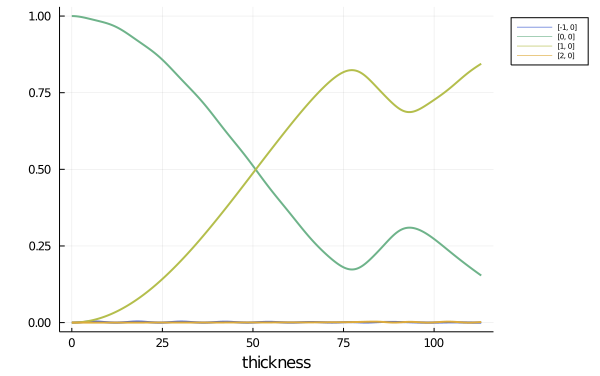}
          \caption{System $(A,F(z_*-z))$}
       \end{subfigure}
       \hfill
        \begin{subfigure}{0.4\textwidth}
           \includegraphics[width=\textwidth]{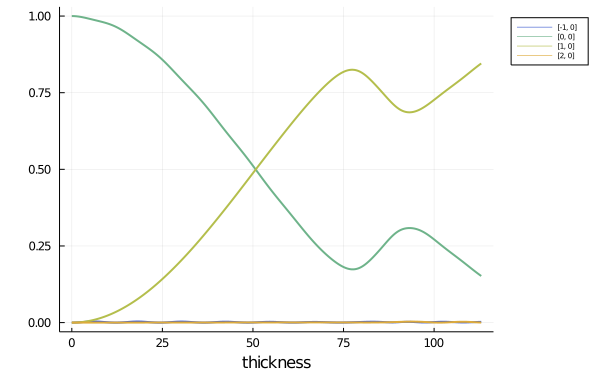}
          \caption{System $(A,-F(z_*-z))$}
       \end{subfigure}
     \caption{Strong and weak symmetry in four-beam approximation: (a) Intensities for system $(A,F(z))$ used as reference
     (b) Intensities for system $(\wt A , \wt F(z)) = (A,-F(z))$. The solution $\wt \phi$ has a strong symmetry compared to the solution $\phi$ of (a).
     (c) Intensities for system $(\hat{A} , \hat{F(z)}) = (A,F(z_*-z))$. The solution $\hat{ \phi}$ has a weak symmetry compared to the solution $\phi$ of (a).
     (d) Intensities for system $(\tilde{A} , \tilde{F(z)}) = (A,-F(z_*-z))$ showing a strong symmetry to case (c) and weak symmetry to (a) and (b).}
     \label{fig:weak_strong_sym}
\end{figure}

    We illustrate strong and weak symmetries by
    numerical simulations of the DHW equations 
    using four beams and a displacement 
    field as given by \eqref{eq:sphere_displacement}, see Figure \ref{fig:weak_strong_sym}. 
    In this example we also observe a dark field symmetry, namely $|\phi_\mathbf{g_{ap}}(z)| = |\wt \phi_\mathbf{g_{ap}}(z)|$ (strong symmetry) or $|\phi_\mathbf{g_{ap}}(z_*)| = |\wt \phi_\mathbf{g_{ap}}(z_*)|$ (weak symmetry) with $\mathbf{g_{ap}}=(1,0)$.
    In Table \ref{4beams} we see the intensities for all four beams at the exit plane.
While Figure \ref{fig:weak_strong_sym} suggests an exact symmetry, Table \ref{4beams} reveals that the symmetry is only \emph{approximate} with an error up to $1\%$. The reason is that the four-beam model does not enjoy the symmetries, however the solutions stay close to the solutions of the two-beam model, see Table \ref{2beams} which has the desired symmetries.
 This simple example demonstrates the importance of the two-beam approximation in the study of symmetries for both bright field and dark field.

\begin{table}
\begin{center}
\begin{tabular}{ |c|c|c|c|c| } 
 \hline
 $\bfg$ & $(A,F(z))$ & $(A,-F(z))$ & $(A,F(z_*-z))$ & $(A,-F(z_*-z))$ \\ \hline
 $(-1,0)$ &0.00012040357\IG{717535638} & \CO{0.00}330035539\IG{26043644}  & \CO{0.000}04461419\IG{606282912}& \CO{0.00}230899563\IG{05712162} \\ 
 $(0,0)$ &  0.15359073146\IG{692287} & \CO{0.15}209371434\IG{729713} & \CO{0.15359073146}\IG{64655}  & \CO{0.15}209371434\IG{861338} \\ 
 $(1,0)$ &  0.84539398729\IG{70054} & \CO{0.84}447759832\IG{29915}  & \CO{0.84}447759832\IG{18772} & \CO{0.84539398729}\IG{72596} \\
 $(2,0)$ &  0.00089487769\IG{78946589} &  \CO{0.000}12833195\IG{71015632} & \CO{0.00}188705604\IG{55906446} & \CO{0.000}20330274\IG{35514181}  \\
 \hline
\end{tabular}
\end{center}
\caption{Comparison of intensities at the exit plane
for the four-beam model in Figure \ref{fig:weak_strong_sym}.
For both bright field ($\mathbf{g}= (0,0)$) and dark field ($\mathbf{g}= (1,0)$) we observe an \emph{approximate symmetry} with an error of about $1\%$.}
\label{4beams}
\end{table}

\begin{table}
\begin{center}
\begin{tabular}{ |c|c|c|c|c| } 
 \hline
 $\bfg$ & $(A,F(z))$ & $(A,-F(z))$ & $(A,F(z_*-z))$& $(A,-F(z_*-z))$ \\ \hline
 $(0,0)$ & 0.15309988945\IG{705253}& \CO{0.15309988945}\IG{87316} &\CO{0.15309988945}\IG{738418} & \CO{0.15309988945}\IG{87582} \\ 
 $(1,0)$ &0.84690011055\IG{19459}  & \CO{0.84690011055}\IG{12671}  & \CO{0.84690011055}\IG{15636} & \CO{0.84690011055}\IG{12052}  \\
 \hline
\end{tabular}
\end{center}
 \caption{Comparison of intensities at the exit plane for the systems in Figure \ref{fig:weak_strong_sym} and under two-beam approximation. Both bright and dark field show a \emph{perfect symmetry} in this case (up to some numerical error).}
 \label{2beams}
 \end{table}

\subsubsection{Three important symmetry facts}
\label{suu:PhysSymmProved}

Here we give an overview of the necessary results from Section \ref{se:Symmetries} that help us explain the symmetries in TEM imaging observed at the beginning of the section. 
The results are stated as facts and put into physics words, while the formal version of them and the proofs can be found in the next section.

The first fact concerns the change in the sign of the strain, which corresponds to changing the sign of $F(z)$,
and is proved in Corollary \ref{co:SignChange}.
\begin{fact}\label{f:sign}
In the two-beam approximation $\DLL_2=\{\mathbf{0},\mathbf{g'}\}$ and under strong beam conditions, i.e. $s_\mathbf{0}=s_\mathbf{g'}=0$, changing the sign of the strain ($F(z) \leadsto -F(z) $) is a strong symmetry.
\end{fact}
 
The next fact concerns reflections at the midplane of the specimen given by the transformation $F(z) \leadsto F(z_*{-}z)$
and is proved in Corollary \ref{co:Flip} part (W3).
\begin{fact}\label{f:midplane}
In the two-beam approximation $\DLL_2=\{\mathbf{0},\mathbf{g'}\}$ a midplane reflection of the strain ($F(z) \leadsto F(z_*{-}z)$) is a weak symmetry.
\end{fact}
Here it is important to notice that Fact \ref{f:midplane} does not require strong beam conditions, so it can be applied for excitation errors $s_\mathbf{g'} \neq 0$.
This result is equivalent to the Type II symmetry in \cite{PT68} or to \cite{HowWhe61DCEM} who showed this symmetry for bright field images.
In the general $m$-beam case the midplane reflection symmetry holds under the assumption that all relevant $U_\bfg$ are real, see part (W2) of Corollary \ref{co:Flip}.

In the next fact we combine the first two facts with an additional sign change of the excitation error $s_\mathbf{g'}$, proved in Corollary \ref{co:ExcErr}. 
\begin{fact}\label{f:exc_err}
In the two-beam approximation $\DLL_2=\{\mathbf{0},\mathbf{g'}\}$ combining the sign change of the strain with a midplane reflection ($F(z) \leadsto -F(z_*-z)$) and changing the sign of the excitation error $s_\mathbf{g'} \leadsto -s_\mathbf{g'}$ is a weak symmetry.
\end{fact}
The Type I symmetry in \cite{PT68} is a special case of this results for $s_{\mathbf{g'}}=0$.
All results are derived for a general strain profile. 
The strain profiles in the examples 
we discussed before have an additional symmetry, 
namely they are even or odd functions which 
are shifted relative to the center of the specimen,
see Figures \ref{fig:quantum_well}d) and \ref{fig:quantum_dot}d).
In the next subsection we will show how the above observations interact with the parity of the strain profile $z \mapsto F(z)$.

\subsubsection{Explanation of observed symmetries}

With the symmetries that we have in hand now we are able to answer all the questions that occurred from the observations we made before.
We start with the symmetry with respect to the sign of the strain ($F(z) \leadsto -F(z)$), that was discussed in Section \ref{su:obs_sign_dis} using the example of the pyramidal quantum dot in Figure \ref{fig:sim_pyramid}. 
We can now say that this is a direct application of 
Fact \ref{f:sign} to every pair of
pixels $(i,j)$ and $(i',j')$ such that $F(z; x_i, x_j) = - F(z; x_{i'}, y_{i'})$ and $F(z)$ being a general strain profile.

\begin{figure}
    \centering
    \includegraphics[width=0.8\textwidth]{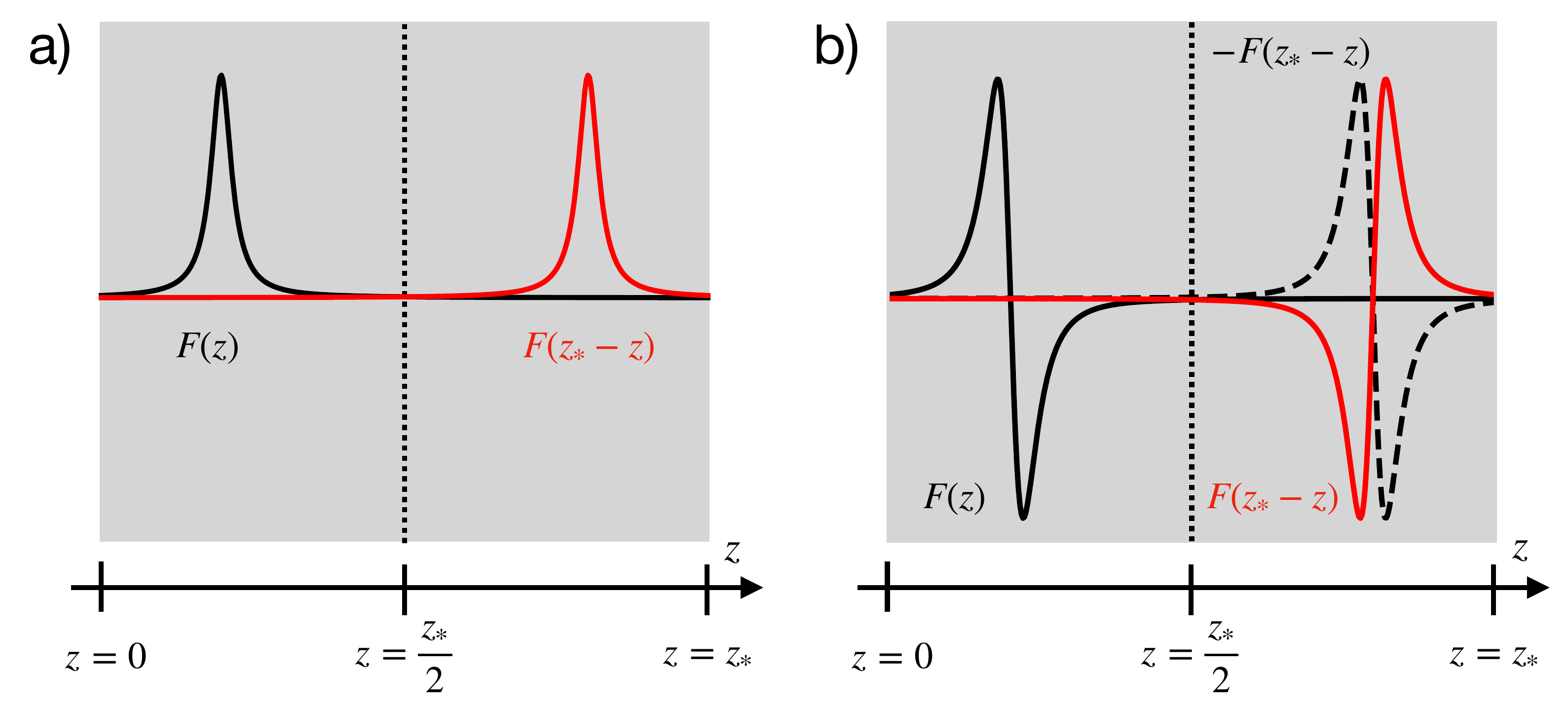}
    \caption{Plot of a shifted even function $F(z)$ (black) and the midplane reflection of it $F(z_*-z)$ (red) illustrating that the midplane reflection corresponds to shifting $F(z)$ a). Plot of a shifted odd function $F(z)$ (black) and the midplane reflection of it (red) illustrating that shifting (black dotted) needs an additional sign change to correspond to midplane reflection b).}
    \label{fig:strain_profiles}
\end{figure}

For the symmetry with respect to the center of the sample discussed in Section \ref{su:obs_shift} a combination of the Facts \ref{f:sign} and \ref{f:midplane} with the parity of the strain profile can explain the observations. 
We take each case separately. 
For the inclined quantum well the strain has an even profile, as in Figure \ref{fig:strain_profiles} a). From Fact \ref{f:midplane} we know that we can apply midplane reflection ($F(z) \leadsto F(z_*-z)$) and get the same pixel intensity. For an even profile midplane reflection and shifting coincide, see Figure \ref{fig:strain_profiles} a). This is the reason why the image shows a pixelwise symmetry with respect to shifting.
In the case of the spherical quantum dot the strain has an odd profile, as in Figure \ref{fig:strain_profiles} b). Applying midplane reflection we don't get the same result as shifting, see Figure \ref{fig:strain_profiles} b). We would need to apply the sign change as well, as stated in Fact \ref{f:sign}. This however can not be done unless we have strong beam conditions (meaning $s_\mathbf{g_{ap}}=0$).   This is the reason why, for $s_\mathbf{g_{ap}} =0 $, we observe a symmetry with respect to shifting while for $s_\mathbf{g_ap} \neq 0$ we don't. 

The observations concerning the sign change of the $s_{\bfg}$ made in Section \ref{su:obs_sign_err},
e.g. see green and red boxes in Figure \ref{fig:qd_spherical_intensities}, 
can be explained from Fact \ref{f:exc_err}:
it says that a midplane reflection combined with a sign change in the strain ($F(z) \leadsto -F(z_*{-}z)$ ) is a symmetry if we also change the sign of the excitation error ($\Sigma \leadsto -\Sigma$).
In this case strong beam condition ($s_\mathbf{g_{ap}} =0$) is not a requirement. 
This means that we can apply midplane reflection plus sign change of the strain, which for the odd strain profile in Figure \ref{fig:qd_spherical_intensities} would correspond to shifting the strain profile with respect to the center, and then change the sign of the excitation error. This explains the symmetric images in Figure \ref{fig:qd_spherical_intensities} indicated by the green and red boxes. The images in Figure \ref{fig:qd_spherical_intensities} indicated by the blue boxes can also be explained now but we will do this in the next section, since they are not pixelwise symmetric as the previous examples but they have a mirror like symmetry.

\subsection{Mirrored TEM images induced by strain}

\begin{figure}
    \centering
    \includegraphics[width=\textwidth]{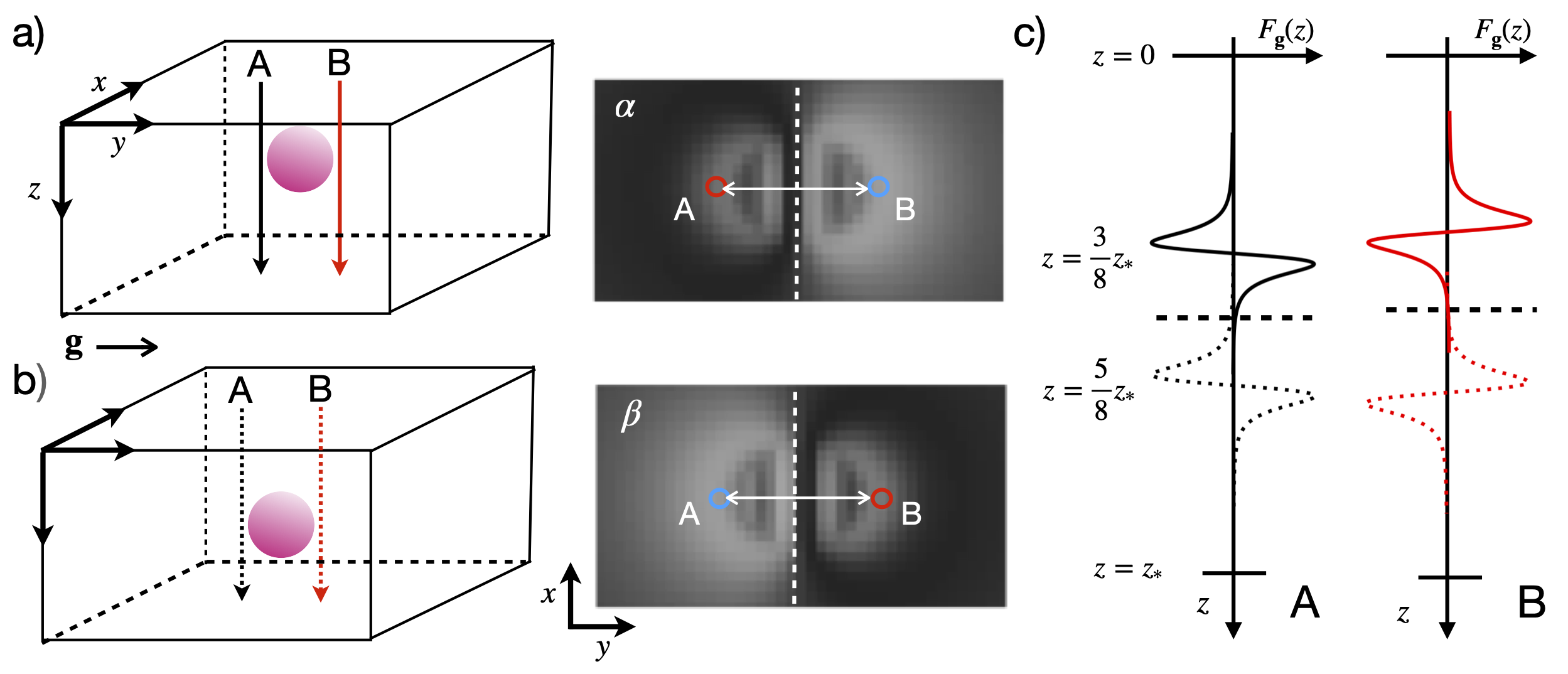}
    \caption{Mirrored images: a) Spherical quantum dot positioned at $z=\fr{3}{8}z_*$ and the corresponding TEM image ($\alpha$). Solving the equations across the line scans A and B gives the corresponding pixels in $\alpha$, denoted also by A and B.
    b) Spherical quantum dot positioned at $z=\fr{5}{8}z_*$ and the corresponding TEM image ($\beta$), showing again the pixels A and B that correspond to the same line scans.
    c) Strain profiles across the line scans A and B. The solid black and red profiles correspond to the TEM image $\alpha$, while the dotted black and red profiles to TEM image $\beta$. The TEM images are adapted from 
    \cite[Fig.5.12]{NiermannPhD21} used under CC-BY, cf. Figure \ref{fig:qd_spherical_intensities}. }
    \label{fig:mirror_sym}
\end{figure}

Here we will focus on explaining the images in Figure \ref{fig:qd_spherical_intensities} that are indicated by the blue boxes. First we start with the TEM images $\alpha$ and $\beta$, see also Figure \ref{fig:mirror_sym}. 
This means we have two images of a spherical quantum dot using the same excitation error, here $s_\mathbf{g_{ap}}=12 \fr{1}{\mu m}$, but placed in different positions, symmetrical to the center of the sample, see Figure \ref{fig:mirror_sym} a) and b).

To analyze the images pixelwise we make two line scans in the $z$ direction, A and B. 
The corresponding pixels for each image are indicated in Figure \ref{fig:mirror_sym} a)(image $\alpha$) and b) (image $\beta$), using the same notation A and B. 
We can see in Figure \ref{fig:mirror_sym} a) that the pixel intensities corresponding to the line scans A and B in image $\alpha$ are not the same. 
So the image itself does not have pixelwise symmetry. 
Comparing the pixels between the images $\alpha$ and $\beta$ though shows that the pixel in the image $\alpha$ that corresponds to the line scan A (or B) is the same as the pixel in the image $\beta$ that corresponds to the line scan B (or A).

To understand these properties using the theory we developed we study the strain profile for each line scan, shown in Figure \ref{fig:mirror_sym} c). First we focus on why the image itself is not pixelwise symmetric. For the quantum dot in image $\alpha$ the strain profile across the two line scans is shown in Figure \ref{fig:mirror_sym} c) by the solid black and red lines. We see that the difference between these two profiles is the sign. Changing the sign of the strain though is a symmetry only under strong beam conditions (Fact \ref{f:sign}) but in this case we have $s_\mathbf{g_{ap}} \neq 0$. The same exact argument applies to image $\beta$.

Next, we compare the two images with each other. 
The strain profile for the spherical quantum dot in image $\beta$ is given in Figure \ref{fig:mirror_sym} c) by the dotted black and red lines. 
The reason that the pixel corresponding to 
the line scan A in image $\alpha$ is equal to 
the one that corresponds to line scan B 
in image $\beta$ is Fact \ref{f:midplane}, since the strain profile for the first case (solid black line in Figure \ref{fig:mirror_sym} c)) is a midplane reflection of the strain profile in the second case (dotted red line in Figure \ref{fig:mirror_sym} c)). This is due to the fact that for an odd function shifting the strain (black dotted line) plus sign change correspond to midplane reflection, see also Figure \ref{fig:strain_profiles}b).

Additionally, from Fact \ref{f:exc_err}, we know that image $\beta$ is symmetric to the image $\gamma$  in Figure \ref{fig:qd_spherical_intensities}. Combining all the above we can see why also the images corresponding to the same position but with opposite excitation errors are mirrored images of each other, see Figure \ref{fig:qd_spherical_intensities}. This mirror-like symmetry is induced by the parity of the strain profile.

\subsection{Symmetries for general profiles} \label{suu:general_prof}

\begin{figure}
    \centering
    \includegraphics[width= \textwidth]{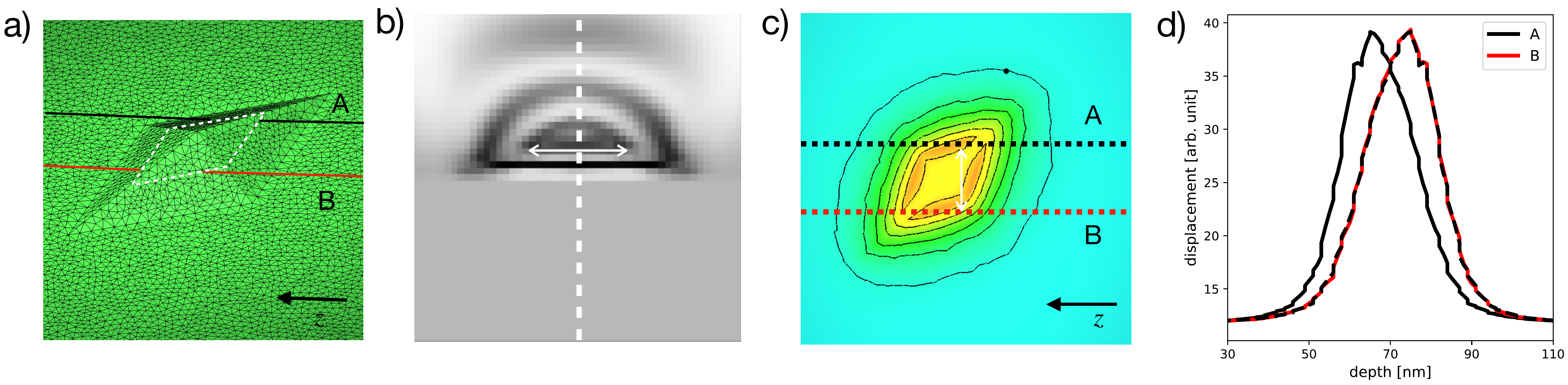}
    \caption{Pyramidal quantum dot with a rhomboidal base : a) 3D geometry showing the $z$ direction and two line scans A and B.
    b) Simulated TEM image for $\mathbf{g_{ap}}= (004)$ showing a pixelwise symmetry.
    c) Displacement component responsible for the 
image contrast at the cut (white dotted lines) in a) and the line scans A and B.
    d) Displacement profiles across the line scans A (black solid) and B (red solid). We see that the displacement profile is not even nor odd and that the displacement in B is the midplane reflection of the displacement in A (black dashed).}
    \label{fig:qd-lat-aspect1}
\end{figure}

The examples discussed until now were for a strain profile 
with odd or even parity. 
However, this parity is not the essential cause for the
symmetries observed between two pixels. 
What is important is the symmetry between 
the strain profiles with respect to sign change 
and midplane reflection. 
To make this clear we consider the case of 
a general strain profile without a specific parity. 
For this purpose we examine TEM images of 
a pyramidal quantum dot with a rhomboid as a base instead
of a square. 
We assume that the quantum dot is
placed at the center of the sample. 
To create these TEM images we used the computational method described 
in \cite{MNSTK20NSTI} and the tool chain employed therein.
First a 3D mesh is generated to represent 
the geometry of the quantum dot using TetGen \cite{tetgen}, see Figure \ref{fig:qd-lat-aspect1} a). 
Then the generated mesh enters the FEM based solver WIAS-pdelib \cite{pdelib}, in order to find the displacement $\bfu$, see Figure \ref{fig:qd-lat-aspect1} c). 
Finally the relevant displacement component enters the DHW solver PyTEM \cite{Nier19pyTEM} in order to simulate the corresponding TEM image, see Figure \ref{fig:qd-lat-aspect1} b).
For this set up two TEM images are computed, corresponding to different vectors $\mathbf{g_{ap}}$ using strong beam excitation conditions.

For an excitation corresponding to $\mathbf{g_{ap}} = (004)$ the TEM image is shown in  \ref{fig:qd-lat-aspect1} b). 
The (projected) component of the displacement, 
which is responsible for the 
image contrast in this case, is shown in \ref{fig:qd-lat-aspect1} c). 
This was taken in a cross-section parallel to the base of the pyramid,
as indicated by the white dotted lines in \ref{fig:qd-lat-aspect1} a).
Next we analyze the displacement profile along the two line scans A and B 
evolving in $z$-direction, as indicated 
in Figures \ref{fig:qd-lat-aspect1} a) and c)
by the black and red dotted lines.
The displacement profiles across these line scans are shown in \ref{fig:qd-lat-aspect1} d), 
where we can see that they are not even or odd. 
However, we observe a pixelwise symmetry in the TEM image.
The displacement $u_A(z)$ across line A is a midplane reflection of the displacement $u_B(z)$ across line B: $u_A(z_*-z) = u_B(z)$. This means that the strain across A differs with the strain across B by a sign plus midplane reflection, $\fr{d}{d z} u_B (z) = - \fr{d}{d z} u_A(z_*-z) $. Then the symmetry we observe in the TEM image follows from Fact \ref{f:exc_err} and due to the strong beam condition ($s_{\bfg_{ap}}=0$).

\begin{figure}
    \centering
    \includegraphics[width=\textwidth]{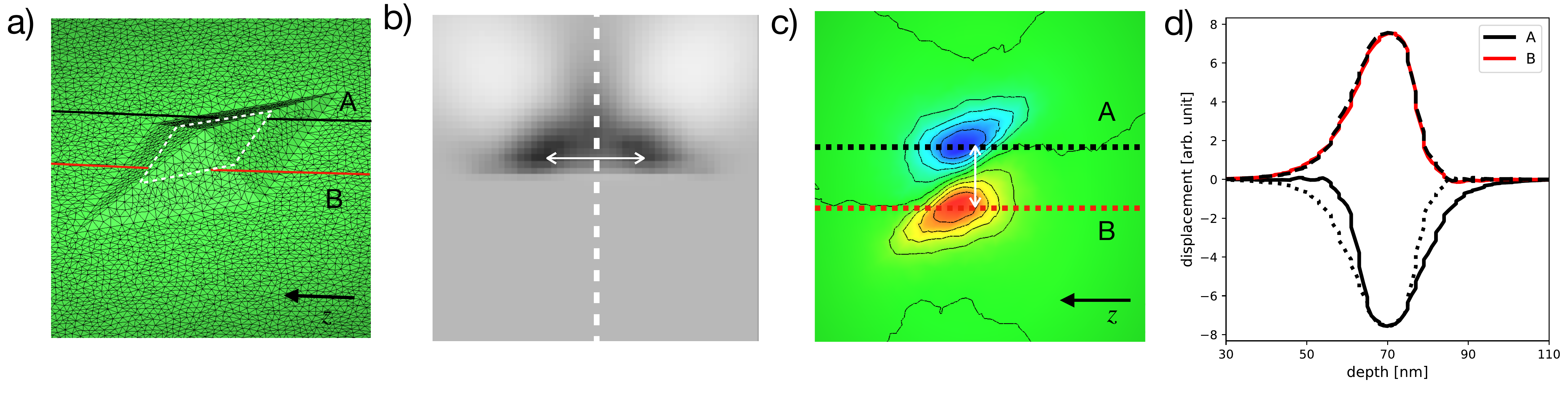}
    \caption{Pyramidal quantum dot with different lateral aspect ratio: a) 3D geometry showing the $z$ direction and the two line scans A and B.
    b) Simulated TEM image for $\mathbf{g_{ap}}= (040)$ showing a pixelwise symmetry.
    c) Displacement component responsible for the 
image contrast at the cut (white dotted lines) 
in a) and the line scans A and B.
    d) Displacement profiles across the line scans A (black solid) and B (red solid). We see that the displacement profile is not even nor odd and that the displacement in B is the midplane reflection (black dotted) plus sign change of the strain in A.}
    \label{fig:qd-lat-aspect2}
\end{figure}

For an excitation corresponding to $\mathbf{g_{ap}} = (040)$ the TEM image is shown in  \ref{fig:qd-lat-aspect2} b). 
Under this excitation the imaging is sensitive to a different component of the displacement field as in the case before.
The corresponding displacement field in the cross-section is shown in \ref{fig:qd-lat-aspect2} c).
We can see that the displacement obeys the sign change symmetry
with respect to the center of the structure, as also observed
for the pyramidal quantum dot with square base, see Figure \ref{fig:sim_pyramid}.
As in the example before, we plot the displacement profile across 
lines A and B as shown in \ref{fig:qd-lat-aspect2} d).
Here again we see that it is not even or odd. However,
midplane reflection and sign change of the displacement profile in A 
equals the displacement in B: $u_B(z) = - u_A(z_*-z)$. This gives for the corresponding strain that $\fr{d}{d z}u_B(z) = \fr{d}{dz}u_A(z_*-z)$. Then Fact \ref{f:midplane} explains the pixelwise symmetry we observe.
These two examples demonstrate that the results from Section \ref{suu:PhysSymmProved} are indeed valid for a general displacement profile.

\section{Mathematical treatment of the symmetries}
\label{se:Symmetries}

We now provide the mathematics underlying the symmetry considerations for the solutions of the DHW equations. For this we use the general $m$-beam model in the Hermitian form derived in \eqref{eq:m-beamHerm}.
To study the symmetries, we consider the system matrix $A=V+\Sigma$ and the strain function $F$ as data specified to lie in the following spaces 
\[
V\in \C^{m\ti m}_\mafo{herm}, \quad \Sigma \in \bbD_m:=\R^{m\ti
  m}_\mafo{diag}, \quad F\in \rmC^0([0,z_*];\bbD_m).
\]

The typical measurements for generating TEM images does not involve all
components of $\phi(z_*)\in \C^m$ at the exit plane, but only the intensity of beam $\bfg_\mafo{ap} \in \DLL_m$ selected by the objective aperture, see Figure \ref{fig:image_formation}a), namely 
$I_\mafo{ap}(x_i,y_j)= |\phi_{\bfg_\mafo{ap}}(z_*;x_i,y_j)|^2$. 
As mentioned above a special mathematical role plays the so-called \emph{bright field} which is given by the choice $\bfg_\mafo{ap}=\mathbf{0}$. The reason for this is the double appearance of the vector $e_\mathbf{0}$, namely (i) in the initial condition $\phi(0)=e_\mathbf{0}$  and (ii) in the exit measurement $\phi_{\bfg_\mafo{ap}}(z_*)= \phi_\mathbf{0}(z_*) = \phi(z_*)\cdot e_\mathbf{0}$. 

The double appearance of $e_\mathbf{0}$ can even be used for symmetries 
in the \emph{dark field} where $\bfg_\mafo{ap}\neq \mathbf{0}$ under 
the assumption that we have a two-beam model, i.e.\ $\DLL_2 = 
\{ \mathbf{0}, \mathbf{g'} \}$ and $\bfg_\mafo{ap}= \mathbf{g'}$. 
In this case we can exploit the Hermitian structure of 
\eqref{eq:m-beamHerm} which provides the simple conservation of 
the Euclidean norm, namely $|\phi(z)| = |\phi(0)|=1$ for all $z\in [0,z_*]$. This property was first derived in \cite[Sec.\,3.1]{KMM21}, where it was related to a wave-flux conservation in the Schr\"odinger equation. In this case we have $ |\phi_\mathbf{g'}(z)|^2 =
1-|\phi_\mathbf{0}(z)|^2$ for all $z\in [0,z_*]$. Thus, if $|\phi_\mathbf{0}(z_*)|$ is preserved by a symmetry, 
then so is $|\phi_\mathbf{g'}(z_*)|$.\medskip 

In light of the above discussions, we are
interested in the question whether 
\medskip

\textbullet\ (sign change) flipping the function $F$ into $-F$ or
\medskip

\textbullet\ (midplane reflection) flipping $F(\;\!\cdot\!\;)$ into $F(z_*{-}\!\;\cdot\!\;)$ 
\\[0.3em]
lead to the same value of $|\phi_\mathbf{0}(z_*)|$ or not. 
\bigskip

To analyze these two symmetries and their joint effect for both, 
$m$-beam models and the two-beam model, we consider more 
general classes of transformations involving also changes of $A=V+\Sigma$ and not only of the strain related part $F$. This will uncover the
proper mathematical structure of the symmetries and show 
why the case $m=2$ is special.  For
this we define two types of symmetries.

\begin{definition}[Strong and weak symmetries]\label{def:Symmetries}
We say that replacing the pair $(A,F)\in \C^{m\ti m}_\mafo{herm} \ti
\rmC^0([0,z_*];\bbD_m)$ by the pair $(\wt A,\wt F)$ is a 
{\bfseries strong symmetry} if the corresponding solutions 
$\phi$ and $\wt\phi$ of \eqref{eq:m-beam} satisfy $|\phi_\mathbf{0}(z)|=|\wt\phi_\mathbf{0}(z)|$ for all $z\in [0,z_*]$. 

We call the replacement a {\bfseries weak symmetry} if we have
$|\phi_\mathbf{0}(z_*)|=|\wt\phi_\mathbf{0}(z_*)|$.  
\end{definition}    

Throughout we will denote by $\bbU_{A+F}(z)\in \C^{m\ti m}$ the evolution operator solving
\[
\dot \bbU = \ii \big(A{+}F(z) \big) \bbU,\qquad \bbU(0)=I.
\]
As $A+F(z)$ is Hermitian for all $z$, the evolution operators $\bbU_{A+F}$
are unitary, i.e.
\begin{equation}
  \label{eq:Unitary}
  \bbU_{A+F}(z)^{-1} = \big(\bbU_{A+F}(z)\big)^* = \ol{\bbU_{A+F}(z)}^\top.
\end{equation}
In particular, this implies that the Euclidean norm $|\phi| = \big(\sum_{\mathbf{g}\in\DLL_m}|\phi_\bfg|^2\big)^{1/2}$ is preserved for solutions $\phi(z)$ of \eqref{eq:m-beamHerm}. 
Of course, we have a general transformation rule for arbitrary unitary matrices $\bbQ \in \C^{m\ti m}$ (i.e.\ $\bbQ^*\bbQ=I$), namely 
\begin{equation}
  \label{eq:UnitaryTrafo}
 \bbU_{\bbQ(A{+}F)\bbQ^*}(z)= \bbQ\, \bbU_{A+F}(z) \bbQ^*.  
\end{equation}

The first result concerns the set of all strong symmetries. 

\begin{proposition}[Strong symmetries]\label{pr:StrongSym}
Any of the following transformations and any composition of them are strong
symmetries:\smallskip 

\emph{(S1)}  simultaneous linear phase factor: $(\wt A,\wt F) = (A{+}\delta I,F)$ 
\smallskip

\emph{(S2)} complex conjugation: $(\wt A,\wt F) = (-\ol A,-F)$
\smallskip

\emph{(S3)} constant phase factors: $(\wt A,\wt F) = (Q_\psi A Q^*_\psi,F)$ with
$Q_\psi=\mafo{diag}(1,\ee^{\ii\psi_2},...,\ee^{\ii \psi_m})$,
\\[0.3em]
where $\delta, \psi_j\in \R$.
\end{proposition}
\begin{proof} In all three cases the result follows easily by writing down the
  corresponding evolution operators.

(S1) $\bbU_{\delta I+A+F}(z) = \ee^{\ii \delta z}\bbU_{A+F}(z)$ giving
$\wt\phi_0(z)=  \ee^{\ii \delta z}\phi_0(z)$. 

(S2) By complex conjugation of \eqref{eq:m-beam} we easily obtain 
$\bbU_{-\ol A-F}(z)=\ol{\bbU_{A+F}(z)}$. As the initial condition $\phi(0)=e_\mathbf{0}$
is real, we conclude $\wt\phi(z)=\ol{\phi(z)}$ and hence
$\wt\phi_\mathbf{0}(z)=\ol{\phi_\mathbf{0}(z)}$. 

(S3) For this case we use the transformation rule \eqref{eq:UnitaryTrafo} 
with $\bbQ=Q_\psi$ and observe that $Q_\psi F(z)Q_\psi^* =F(z)$ because $F$ is
diagonal. Hence we have $\wt\phi_\mathbf{0}(z)=\phi_\mathbf{0}(z)$.
\end{proof}

As a first nontrivial result we now reduce to the case $m=2$ with the
additional restriction $A_{\mathbf{0}\mathbf{0}}=A_{\mathbf{g'}\mathbf{g'}}$. Indeed, the condition 
\[
 A_{\mathbf{00}}=A_{\mathbf{g'g'}},  \quad \text{ which  means }
 \frac{U_\mathbf{0} } {\rho_\mathbf{0} } + 2 s_\mathbf{0} = 
 \frac{U_\mathbf{0} } {\rho_\mathbf{g'} } + 2 s_\mathbf{g'},
\]
is typically satisfied (in high enough accuracy) in the case of the strong two-beam conditions, 
because one usually chooses $s_\mathbf{g'}=s_\mathbf{0}=0$ and one has $\rho_\mathbf{0} = \mathbf{k_0} \cdot \bfnu \approx \rho_\mathbf{g'} $. This holds automatically if $\mathbf{g'}\cdot \bfnu=0$ or it is approximately true in the case of high energy electrons, i.e.~$|\mathbf{k_0}| \gg  |\mathbf{g'}|$. 

\begin{corollary}[Sign change using $m=2$ and $A_{\mathbf{0}\mathbf{0}}=A_{\mathbf{g'}\mathbf{g'}}$]
\label{co:SignChange}
In the case $A=V+\Sigma \in \C^{2\ti 2}_\mafo{Herm}$ with $A_{\mathbf{0}\mathbf{0}}=A_{\mathbf{g'}\mathbf{g'}}$, the
transformation $(\wt A,\wt F)= (A,-F)$ is a strong symmetry, i.e.\
$|\wt\phi_\bfg(z)|= | \phi_\bfg(z)|$ for $z \in [0, z_*]$ and $\bfg \in \DLL_2= \{ \mathbf{0},\mathbf{g'} \}$.
\end{corollary}
\begin{proof} The result follows by combining the three strong symmetries
  (S1)--(S3). We write 
\[
A=\bma{cc} a& b\\ \ol b& a \ema \quad \text{with } a\in \R \text{ and }
b= |b|\ee^{\ii \beta}.
\]
Applying first (S2) we find a strong symmetry with 
$(A_1,F_1)=(-\ol A,-F)$. Next we apply (S1) with $\delta=2 a$ such 
that $(A_2,F_2)=(2aI-\ol A,-F)$ is again a strong symmetry. 
Finally we apply (S3) with $\psi_2=\pi-2\beta$ and observe
that $ \ee^{\ii \psi_2}= - \ee^{-\ii 2\beta}$, which gives  
\[
  \mafo{diag}(1,-\ee^{\ii 2\beta}) \,\big( 2 a I {-} \ol A\big)\,  \mafo{diag}(1, -\ee^{-\ii 2\beta}) = A.
\]
Hence, $(A_3,F_3)=(A,-F)$ is a strong symmetry giving
$|\wt\phi_\mathbf{0}(z)|=|\phi_\mathbf{0}(z)|$ for all $z \in [0,z_*]$. 

Finally, the assumption $m=2$ and the unitarity \eqref{eq:Unitary} 
give, for $\eta=\phi$ or $\wt\phi$, the relation 
\[
|\eta_\mathbf{0}(z)|^2 \overset{m=2}= |\eta(z)|^2 -|\eta_\mathbf{g'}(z)|^2 \overset{\text{unit.}}=  |e_\mathbf{0}|^2-|\eta_\mathbf{g'}(z)|^2. 
\]
Hence, we obtain $|\wt\phi_\mathbf{g'}(z)| =|\phi_\mathbf{g'}(z)|$ from the corresponding result for $\bfg=\mathbf{0}$.   
\end{proof} 

To study the midplane reflection we introduce the \[
\text{\em flip operator} \quad R(z)=z_*{-}z
\]
acting on $\rmC^0([0,z_*];\bbD_m)$ via
$(F{\circ}R)(z)=F(R(z))=F(z_*{-}z)$. The following identity will 
be crucial for the understanding of the flip symmetry
as a weak symmetry. Of course, one cannot expect that flipping gives rise to a strong symmetry. To see this we consider a nontrivial strain profile $F$ with $F(z)=0$ for $z\in [z_*/2,z_*]$, i.e.\ the perturbation acts only in the upper half of the specimen. The flipped case $\wt F= F {\circ} R$ then corresponds to a perturbation acting only in the lower half of the specimen. In such a case one cannot expect that the bright-field intensities $|\phi_\mathbf{0}(z)|^2$ and 
$|\wt\phi_\mathbf{0}(z)|^2$ are the same inside the specimen. However, because of the double occurrence of the vector $e_\mathbf{0}$ there is some chance that the intensities match for $z=z_*$ only.

\begin{lemma}[Reversal of direction] For all $A \in \C^{m\ti m}_\mafo{herm}$ and $F\in \rmC^0([0,z_*];
\C^{m\ti m}_\mafo{herm})$ we have the identity 
\begin{equation}
  \label{eq:bbU.flip}
 \bbU_{-A-F{\circ}R}(z_*)=\big[\bbU_{A+F}(z_*)\big]^* . 
\end{equation}
\end{lemma} 
\begin{proof}
We set $\wt\bbU(z)=\bbU_{A+F}(z_*{-}z)$, which obviously
satisfies $\wt\bbU(z_*)=I$ and 
\[
\dot{\wt\bbU}(z)=-\dot{\bbU}_{A+F}(z_*{-}z) = -\ii \big(A{+}F(z_*{-}z) \big)
\bbU_{A+F}(z_*{-}z) =  \ii \big( {-}A {-}(F{\circ}R)(z)) \wt\bbU(z).
\]
Thus, $\wt\bbU$ satisfies the same ODE as $\bbU_{-A-F{\circ}R}$, but the initial
conditions are different. This observation, $\wt\bbU(z_*)=I$, and the unitarity
relation \eqref{eq:Unitary} imply  
\[
\bbU_{-A-F{\circ}R}(z)=  \bbU_{A+F}(z_*{-}z) \big[\bbU_{A+F}(z_*)\big]^{-1} 
= \bbU_{A+F}(z_*{-}z) \big[\bbU_{A+F}(z_*)\big]^* .
\]
Restricting to the case $z=z_*$ gives the desired assertion. 
\end{proof}

Of course, all compositions of a strong symmetry with a weak 
symmetry again provides a weak symmetry. Hence, combining the 
above lemma with Proposition \ref{pr:StrongSym} gives the 
following result that relies on the double occurrence of $e_\mathbf{0}$ in the definition of weak symmetries. 

\begin{corollary}[Flipping with $R$ as weak symmetry] 
\label{co:Flip}
For all $A \in \C^{m\ti m}_\mafo{herm}$ and
  $F\in \rmC^0([0,z_*];\bbD_m)$ the following transformations are weak symmetries:

\emph{(W1)} \quad $(\wt A,\wt F) = (-A,-F{\circ} R)$\smallskip 

\emph{(W2)}  \quad $(\wt A,\wt F) = (\ol A,F{\circ} R)$ \smallskip 

\emph{(W3)}   \quad $(\wt A,\wt F) = (A,F{\circ} R)$ in the case $m=2$.\smallskip  
\end{corollary} 
\begin{proof}
We use that weak symmetry is defined in terms of 
\[
\wt\phi_\mathbf{0}(z_*)=\big< \wt\phi(z_*),e_\mathbf{0} \big> = \big< \bbU_{\wt A+\wt F}(z_*)
e_\mathbf{0},e_\mathbf{0} \big> ,
\]
where $e_\mathbf{0}$ occurs as initial condition as well as test vector at $z=z_*$. 

For (W1) we exploit the relation \eqref{eq:bbU.flip} from the previous lemma, which gives
\[
\wt\phi_\mathbf{0}(z_*)= \big< \bbU_{-A-F\circ R}(z_*) e_\mathbf{0},e_\mathbf{0} \big>= 
\big< \bbU_{A+F}(z_*)^* e_\mathbf{0},e_\mathbf{0} \big>= 
\big< e_\mathbf{0},\bbU_{A+F}(z_*)^* e_\mathbf{0} \big>= \ol{\phi_\mathbf{0}(z_*)}.
\]
This immediately implies $|\wt\phi_\mathbf{0}(z_*)| = |\phi_\mathbf{0}(z_*)|$ as desired. 
For (W2) we simply apply the complex conjugation (S2) and use that 
$F$ is real-valued. 
 
For (W3) we start from (W2) and use $m=2$ to replace $\ol A$ by $A$ using (S3) as for
Corollary \ref{co:SignChange}.%  
\end{proof}

From symmetry (W2) follows that under the assumption that all relevant Fourier coefficients of the scattering potential $U_\mathbf{g}$ are real the midplane reflection symmetry 
is also valid for the general m-beam model and not only 
for the two-beam approximation. 
This property may be satisfied for specifc crystal structures.
One example are centrosymmetric materials, such as Al, Cu, and Au
obeying a face-centered cubic lattice, see 
\cite[Ch. 6.5]{Degr03ICTE}.

Our last result concerns a symmetry in the two-beam model when one changes the sign of the excitation error $s_\mathbf{g'}$. This is relevant in experimental observations, where $s_\mathbf{g'}$ can easily be varied, cf. \cite{NiermannPhD21}. In particular, we refer to the Figures \ref{fig:quantum_well}, \ref{fig:quantum_dot}, and \ref{fig:qd_spherical_intensities}. 

\begin{corollary}[Excitation-error symmetry for $m=2$]
\label{co:ExcErr} 
Consider $\DLL_2=\{\mathbf{0},\mathbf{g'}\}$,
$F\in \rmC^0([0,z_*];\bbD_2)$,  
and $A = V + \Sigma \in \C^{2\ti 2}_\mafo{Herm}$ with $V_{\mathbf{0}\mathbf{0} }=V_{\mathbf{g'}\mathbf{g'}}$. Then, 
the transformation $(\wt A,\wt F) = (V{-}\Sigma,-F{\circ} R)$ is a 
weak symmetry.\smallskip 
\end{corollary}
\begin{proof}
The result follows by combining Corollary \ref{co:SignChange} and part (W3) of Corollary \ref{co:Flip}. 
More precisely, we first observe $
\bbU_{A{+}F} = \bbU_{V+(\Sigma{+}F)}$. Applying Corollary \ref{co:SignChange} with $(A,F)$ replaced by $(V,\Sigma{+}F)$ yields that $(A_1,F_1)=(V,-(\Sigma{+}F))$ is a strong symmetry. Combining this with part (W3) of Corollary \ref{co:Flip} shows that 
$(A_2,F_2)  = (V,-(\Sigma{+}F){\circ}R)$ is a weak symmetry. 

To conclude we observe that $\Sigma{\circ} R = \Sigma$ because 
$\Sigma$ is constant. Moving $-\Sigma$ into $\wt A=V-\Sigma$, we 
see that $(\wt A,\wt F)=(V{-}\Sigma, - F{\circ} R)$  is indeed a weak symmetry. 
\end{proof}

\section{Conclusion}
\label{se:Conclusion}
The symmetry properties of the TEM imaging process were analyzed via the DHW equations. This analysis showed that the imaging process is invariant under special transformations.
The most important symmetries are the sign change 
of the strain field and the midplane reflection as well as a 
symmetry related to the sign change of the excitation error.
The latter can be of particular importance in experiments, since modern transmission electron microscopes can easily create series of images by changing the excitation error.
Combining these results with specific properties of the strain profile of the inclusion explains 
extra symmetries observed in TEM images.
The distinction between symmetries of the imaging process and symmetries of the strain field can be used to extract information for the inclusion, e.g. shape or size. The approach can also be
applied to the imaging of dislocations, since the TEM images 
are sensitive to the strain field they induce.
  
\paragraph*{Acknowledgments.} 
The authors are grateful to Tore
and Laura Niermann  for helpful discussions
and to Timo Streckenbach for his support in creating the 3D graphics, cross-sections and the line scans using WIAS-gltools, which have been used in Figures \ref{fig:qd-lat-aspect1} and \ref{fig:qd-lat-aspect2}.
The research was partially supported by
the DFG via  through
the Berlin Mathematics Research Center MATH+
(EXC-2046/1, project ID: 390685689) via the subproject EF3-1 \emph{Model-based
  geometry reconstruction from TEM images.}

\footnotesize

\addcontentsline{toc}{section}{References}

\bibliographystyle{my_alpha}
\bibliography{DHW_bib_03}

\newcommand{\etalchar}[1]{$^{#1}$}
\begin{thebibliography}{11}\itemsep0.1em

\bibitem[BF{\etalchar{*}}64]{BFTTM64}
{\scshape D.~E.~Bilhorn, L.~L.~Foldy, R.~M.~Thaler, W.~Tobocman, {\upshape and}
  V.~A.~Madsen}.
\newblock Remarks concerning reciprocity in quantum mechanics.
\newblock {\em Journal of Mathematical Physics}, 5(4), 435--441, 1964.

\bibitem[Bra13]{Bra13}
{\scshape W.~L.~Bragg}.
\newblock The structure of some crystals as indicated by their diffraction of
  {X-rays}.
\newblock {\em R. Soc. Lond. A}, 89, 248–277, 1913.

\bibitem[CaI16]{CasIsm2016CP}
{\scshape E.~Castellani {\upshape and} J.~Ismael}.
\newblock Which {Curie}’s principle?
\newblock {\em Philosophy of Science}, 83(5), 2016.

\bibitem[Dar14]{Darw14TXRR12}
{\scshape C.~G.~Darwin}.
\newblock The theory of {X}-ray reflexion. {Part I and II}.
\newblock {\em Phil. Mag.}, 27(158+160), 315--333 and 675--690, 1914.

\bibitem[{De }03]{Degr03ICTE}
{\scshape M.~{De Graf}}.
\newblock {\em Introduction to Conventional Transmission Electron Microscopy}.
\newblock Cambridge University Press, 2003.

\bibitem[Ewa21]{Ewal21BOEG}
{\scshape P.~P.~Ewald}.
\newblock {Die Berechnung optischer und elektrostatischer Gitterpotentiale}.
\newblock {\em Annalen der Physik}, 3, 253--287, 1921.

\bibitem[FS{\etalchar{*}}19]{pdelib}
{\scshape J.~Fuhrmann, T.~Streckenbach, {\upshape and} ~others}.
\newblock pdelib: A finite volume and finite element toolbox for {PDE}s.
  [{S}oftware], 2019.

\bibitem[FT{\etalchar{*}}72]{FTKKU72}
{\scshape F.~Fujimoto, S.~Takagi, K.~Komaki, H.~Koike, {\upshape and}
  Y.~Uchida}.
\newblock The reciprocity of electron diffraction and electron channeling.
\newblock {\em Radiation Effects}, 12(3-4), 153--161, 1972.

\bibitem[HoW61]{HowWhe61DCEM}
{\scshape A.~Howie {\upshape and} M.~J.~Whelan}.
\newblock Diffraction contrast of electron microscope images of crystal lattice
  defects. {II. The} development of a dynamical theory.
\newblock {\em Proc. Royal Soc. London Ser.~A}, 263(1313), 217--237, 1961.

\bibitem[HWM62]{HWM62}
{\scshape A.~Howie, M.~J.~Whelan, {\upshape and} N.~F.~Mott}.
\newblock Diffraction contrast of electron microscope images of crystal lattice
  defects. {III. Results} and experimental confirmation of the dynamical theory
  of dislocation image contrast.
\newblock {\em Proceedings of the Royal Society of London. Series A.
  Mathematical and Physical Sciences}, 267(1329), 206--230, 1962.

\bibitem[ISWS74]{ISW74}
{\scshape J.~C.~Ingram, F.~R.~Strutt, {\upshape and} ~Wen-Shiantzeng}.
\newblock An analysis of the symmetries in electron microscope images of a
  sloping down dislocation and its application as a method for dislocation
  characterization.
\newblock {\em Phys. Status Solidi A}, 22, 1974.

\bibitem[Jam90]{Jame90APTH}
{\scshape R.~James}.
\newblock {\em Applications of perturbation theory in high energy electron
  diffraction}.
\newblock PhD thesis, University of Bath, 1990.

\bibitem[Kat80]{Katerbau80}
{\scshape K.~H.~Katerbau}.
\newblock Diffraction contrast and defect symmetry.
\newblock {\em Physica Status Solidi (a)}, 59, 211--221, 1980.

\bibitem[Kir20]{Kirk20ACEM}
{\scshape E.~J.~Kirkland}.
\newblock {\em Advanced Computing in Electron Microscopy}.
\newblock Springer, 3rd edition, 2020.

\bibitem[KMM21]{KMM21}
{\scshape T.~Koprucki, A.~Maltsi, {\upshape and} A.~Mielke}.
\newblock On the {Darwin--Howie--Whelan} equations for the scattering of fast
  electrons described by the {Schr\"odinger} equation.
\newblock {\em SIAM Journal on Applied Mathematics}, 81(4), 1552--1578, 2021.

\bibitem[MN{\etalchar{*}}19]{MNBL19DDEG}
{\scshape L.~Mei{\ss}ner, T.~Niermann, D.~Berger, {\upshape and} M.~Lehmann}.
\newblock Dynamical diffraction effects on the geometric phase of inhomogeneous
  strain fields.
\newblock {\em Ultramicroscopy}, 207, 112844, 2019.

\bibitem[MN{\etalchar{*}}20]{MNSTK20NSTI}
{\scshape A.~Maltsi, T.~Niermann, T.~Streckenbach, K.~Tabelow, {\upshape and}
  T.~Koprucki}.
\newblock Numerical simulation of tem images for in(ga)as/gaas quantum dots
  with various shapes.
\newblock {\em Opt. Quantum Electr.}, 52, 1--11, 05 2020.

\bibitem[Moo72]{Moodie72}
{\scshape A.~F.~Moodie}.
\newblock Reciprocity and shape functions in multiple scattering diagrams.
\newblock {\em Zeitschrift f\"ur Naturforschung A}, 27(3), 437--440, 1972.

\bibitem[Nie19]{Nier19pyTEM}
{\scshape T.~Niermann}.
\newblock {pyTEM}: A python-based {TEM image} simulation toolkit.
\newblock Software, 2019.

\bibitem[Nie21]{NiermannPhD21}
{\scshape L.~Niermann}.
\newblock {\em Untersuchung und {A}nwendung der dynamischen {B}eugung an
  inhomogenen {V}erschiebungsfeldern in {E}lektronenstrahlrichtung in
  {H}albleiterheterostrukturen}.
\newblock Doctoral thesis, Technische Universität Berlin, Berlin, 2021.

\bibitem[PH{\etalchar{*}}18]{Pascal2018}
{\scshape E.~Pascal, B.~Hourahine, G.~{Naresh-Kumar}, K.~Mingard, {\upshape
  and} C.~{Trager-Cowan}}.
\newblock Dislocation contrast in electron channelling contrast images as
  projections of strain-like components.
\newblock {\em Materials Today: Proceedings}, 5, 14652--14661, 2018.

\bibitem[PoT68]{PT68}
{\scshape A.~P.~Pogany {\upshape and} P.~S.~Turner}.
\newblock {Reciprocity in electron diffraction and microscopy}.
\newblock {\em Acta Crystallographica Section A}, 24(1), 103--109, Jan 1968.

\bibitem[QiG89]{QG89}
{\scshape L.~Qin {\upshape and} P.~Goodman}.
\newblock An alternative study of the reciprocity theorem in electron
  diffraction.
\newblock {\em Ultramicroscopy}, 27(1), 115--116, 1989.

\bibitem[ScS93]{SchSta93}
{\scshape R.~Sch\"aublin {\upshape and} P.~Stadelmann}.
\newblock A method for simulating electron microscope dislocation images.
\newblock {\em Mater.\ Sci.\ Engin.}, A164, 373--378, 1993.

\bibitem[Si15]{tetgen}
{\scshape H.~Si}.
\newblock Tetgen, a delaunay-based quality tetrahedral mesh generator.
\newblock {\em ACM Transactions on Mathematical Software}, 41, 1--36, 2015.

\bibitem[WuS19]{WuSch19TEMD}
{\scshape W.~Wu {\upshape and} R.~Schaeublin}.
\newblock {TEM} diffraction contrast images simulation of dislocations.
\newblock {\em J. Microscopy}, 275(1), 11--23, 2019.

\bibitem[Zh{D}20]{ZhuDeg20EBSD}
{\scshape C.~Zhu {\upshape and} M.~{De Graef}}.
\newblock {EBSD} pattern simulations for an interaction volume containing
  lattice defects.
\newblock {\em Ultramicroscopy}, 218, 113088/1--12, 2020.

\end{thebibliography}

%Authors identification:

%T. K. : \texttt{thomas.koprucki@wias-berlin.de}
%ORCID: 0000-0001-6235-9412

%A. Ma. : \texttt{anieza.maltsi@wias-berlin.de}
%ORCID: 0000-0003-2417-8770

%A. Mi. : \texttt{alexander.mielke@wias-berlin.de}
%ORCID: 0000-0002-4583-3888

\end{document}